\documentclass{osa-article}

\journal{osajournal}

\articletype{Research Article}

\begin{document}

\title{Resolution analysis in a lens-free on-chip digital holographic microscope}

\author{Jialin Zhang,\authormark{1,2,3} Jiasong Sun,\authormark{1,2,3} Qian Chen,\authormark{2,4} and Chao Zuo\authormark{2,3,*}}

\address{\authormark{1}School of Electronic and Optical Engineering, Nanjing University of Science and Technology, No. 200 Xiaolingwei Street, Nanjing, Jiangsu Province 210094, China\\
\authormark{2}Jiangsu Key Laboratory of Spectral Imaging \& Intelligent Sense, Nanjing, Jiangsu Province 210094, China\\
\authormark{3}Smart Computational Imaging Laboratory (SCILab), Nanjing University of Science and Technology, Nanjing, Jiangsu Province 210094, China\\
\authormark{4}chenq@njust.edu.cn}
\email{\authormark{*}zuochao@njust.edu.cn} 



\begin{abstract}
Lens-free on-chip digital holographic microscopy (LFOCDHM) is a modern imaging technique whereby the sample is placed directly onto or very close to the digital sensor, and illuminated by a partially coherent source located far above it. The scattered object wave interferes with the reference (unscattered) wave at the plane where a digital sensor is situated, producing a digital hologram that can be processed in several ways to extract and numerically reconstruct an in-focus image using the back propagation algorithm. Without requiring any lenses and other intermediate optical components, the LFOCDHM has unique advantages of offering a large effective numerical aperture (NA) close to unity across the native wide field-of-view (FOV) of the imaging sensor in a cost-effective and compact design. However, unlike conventional coherent diffraction limited imaging systems, where the limiting aperture is used to define the system performance, typical lens-free microscopes only produce compromised imaging resolution that far below the ideal coherent diffraction limit. At least five major factors may contribute to this limitation, namely, the sample-to-sensor distance, spatial and temporal coherence of the illumination, finite size of the equally spaced sensor pixels, and finite extent of the image sub-FOV used for the reconstruction, which have not been systematically and rigorously explored until now. In this work, we derive five transfer function models that account for all these physical effects and interactions of these models on the imaging resolution of LFOCDHM. We also examine how our theoretical models can be utilized to optimize the optical design or predict the theoretical resolution limit of a given LFOCDHM system. We present a series of simulations and experiments to confirm the validity of our theoretical models.
\end{abstract}

\section{Introduction}
High-throughput optical microscopy is essential to various biomedical applications such as cell cycle assay, drug development, digital pathology, and high-content biological screening \cite{maricq1973patterns,Huisman2010Creation}. For conventional whole slide imaging (WSI) systems, in order to capture a high-throughput image with both high-resolution and large field of view (FOV), mechanical scanning and stitching are required to expand the limited FOV of a conventional high magnification objective \cite{ma2007use}, which not only complicate the imaging process, but also significantly increase the overall cost of the system. The recently developed computational microscopy techniques provide new opportunities to create high-resolution wide FOV images without any mechanical scanning and stitching, such as synthetic aperture interferometric microscopy \cite{mico2006synthetic,yuan2008angular,Hillman2009High,kim2014common,kim2014profiling,lim2015comparative}, Fourier ptychography microscopy (FPM) \cite{Zheng2013Wide,ou2013quantitative,tian2014multiplexed,ou2015high,zuo2016adaptive,sun2016efficient,sun2016sampling}, and lens-free on-chip microscopy \cite{zheng2011epetri,luo2016propagation,rivenson2018deep,zhang2018lensfree}. Among these approaches, the lens-free on-chip microscopy has unique advantages of achieving a large effective numerical aperture (NA) $\sim 1$ across the native FOV of the imaging sensor tens of $mm^2$, based on a so-called unit-magnification configuration, where the samples are placed as close as possible to the imaging sensor \cite{garcia2006immersion,ozcan2016lensless}. Without requiring any lenses and other optical components between the object and the sensor planes, lens-free on-chip microscopy allows to significantly simplify the imaging system and meanwhile effectively circumvent the optical aberrations and chromaticity that are inevitable in conventional lens-based imaging systems \cite{mudanyali2010compact,su2010compact}. There are two typical designs for a lens-free on-chip microscope, so-called contact-mode shadow imaging-based microscope \cite{cui2008lensless,zheng2011epetri} and lens-free on-chip digital holographic microscope (LFOCDHM) \cite{garcia2006immersion,bishara2011holographic}. In the contact-mode shadow imaging-based microscopes, the distance between the sample and the sensor need to be quite small (typically less than 10 $\mu m$), and the captured shadows of the objects can be regarded as a two-dimensional absorption image of the specimen \cite{greenbaum2012imaging}. However, the small distance is very difficult to achieve in practice due to the existence of protective glass covering the surface of the camera sensor. In LFOCDHM, the distance between the objects and the sensor chip can be sizeable, and diffraction patterns are generated from the interference between the scattered light from each object and itself or the unscattered background light. The diffraction patterns are be digitally processed to reconstruct an image of the specimen, and the associated twin-image artifacts need to be eliminated or partially removed relying on computational phase retrieval algorithm \cite{barton1991removing,latychevskaia2007solution}. In the following analysis, we will examine LFOCDHM exclusively.

Despite the advantages mentioned earlier, the LFOCDHM systems generally suffer from low imaging resolution which is far from enough to meet the demand of recent biomedical research, particularly with respect to the visualization of cellular or subcellular details of biological structures and processes. Unlike conventional coherent diffraction limited imaging systems, where the limiting aperture is used to define the system performance, typical LFOCDHM systems only produce compromised imaging resolution that far below the ideal coherent diffraction limit. According to Nyquist-Shannon sampling theorem, the resolution of the holographic reconstruction is fundamentally limited to the sampling resolution of the imaging devices since the recorded holographic fringes are not magnified. In other words, the physical pixel-size is one important limiting factor of these lens-free imaging systems \cite{greenbaum2012imaging}. Because of the spatial aliasing/undersampling, the imaging sensor will fail to record holographic oscillation corresponding to high spatial frequency information of the specimen. To address this problem, pixel super-resolution (SR) methods have been proposed in which the hologram with a smaller effective pixel size can be synthesized from multiple low-resolution (LR) measurements through specific computational algorithms \cite{cui2008lensless,zheng2011epetri,bishara2011holographic,park2003super,luo2016propagation}. With these pixel SR methods, the imaging resolution of the LFOCDHM systems can be improved from Nyquist-Shannon limit (half-pitch lateral resolution of  $\sim 2 \mu m$, effective NA of $\sim 0.1-0.2$) to an effective numerical aperture of $\sim 0.4-0.5 $ \cite{zheng2011epetri,bishara2011holographic,luo2016propagation,zhang2017adaptive}. Even though the achieved imaging resolution is still only less than half of the ideal coherent diffraction limit (NA $\sim 1$). The reason for this is that besides the pixel size of the sensor, at least 4 additional factors act to significantly limit the performance of LFOCDHM systems, namely, the sample-to-sensor distance, spatial and temporal coherence of the illumination, and finite extent of the image sub-FOV used for the reconstruction. This is not unexpected and has been discussed by other authors see, for example, Refs. \cite{greenbaum2012imaging,luo2016propagation}. However, either only qualitative analyses were presented \cite{parrent1965resolution,agbana2017aliasing}, or only one or two of these factors on the imaging resolution have been considered \cite{xu2005imaging,kelly2009resolution,hao2011resolution,agbana2017aliasing}. In these quantitative analyses \cite{xu2005imaging,kelly2009resolution,hao2011resolution}, the discrete features of the sensor attract more attention, but the other basic parameters, e.g., the sample-to-sensor distance \cite{doblas2015study}, spatial and temporal coherence of the illumination \cite{agbana2017aliasing}, and finite extent of the image sub-FOV \cite{kelly2013filtering}, are sporadically mentioned in the off-axis/in-line digital holographic microscopy. Thus, the influence of these 5 factors on the imaging resolution of LFOCDHM has not been systematically examined and rigorously explored until now.

In this work, we have conducted a systematical research on the effect of five major factors on imaging resolution of a LFOCDHM system, i.e., the sample-to-sensor distance, spatial and temporal coherence of the illumination,  finite size of the equally spaced sensor pixels, and finite extent of the image sub-FOV used for the reconstruction. We derive five transfer function models that account for all these physical effects and their interactions on the imaging resolution of LFOCDHM. We further combine all these effects into a unified transfer function, which is the continued multiplication of the five sub-transfer functions. We examine how these theoretical models can be utilized to predict the theoretical resolution limit of a given LFOCDHM system or provide a useful guide to the selection of different system parameters for the optimization of the imaging resolution when designing a new LFOCDHM system. A series of simulations and experiments are presented to confirm the validity of our theoretical models.

\section{Principle}
\subsection{Typical optical setup for LFOCDHM}
\begin{figure}[!b]
\centering
\includegraphics[width=0.75\linewidth]{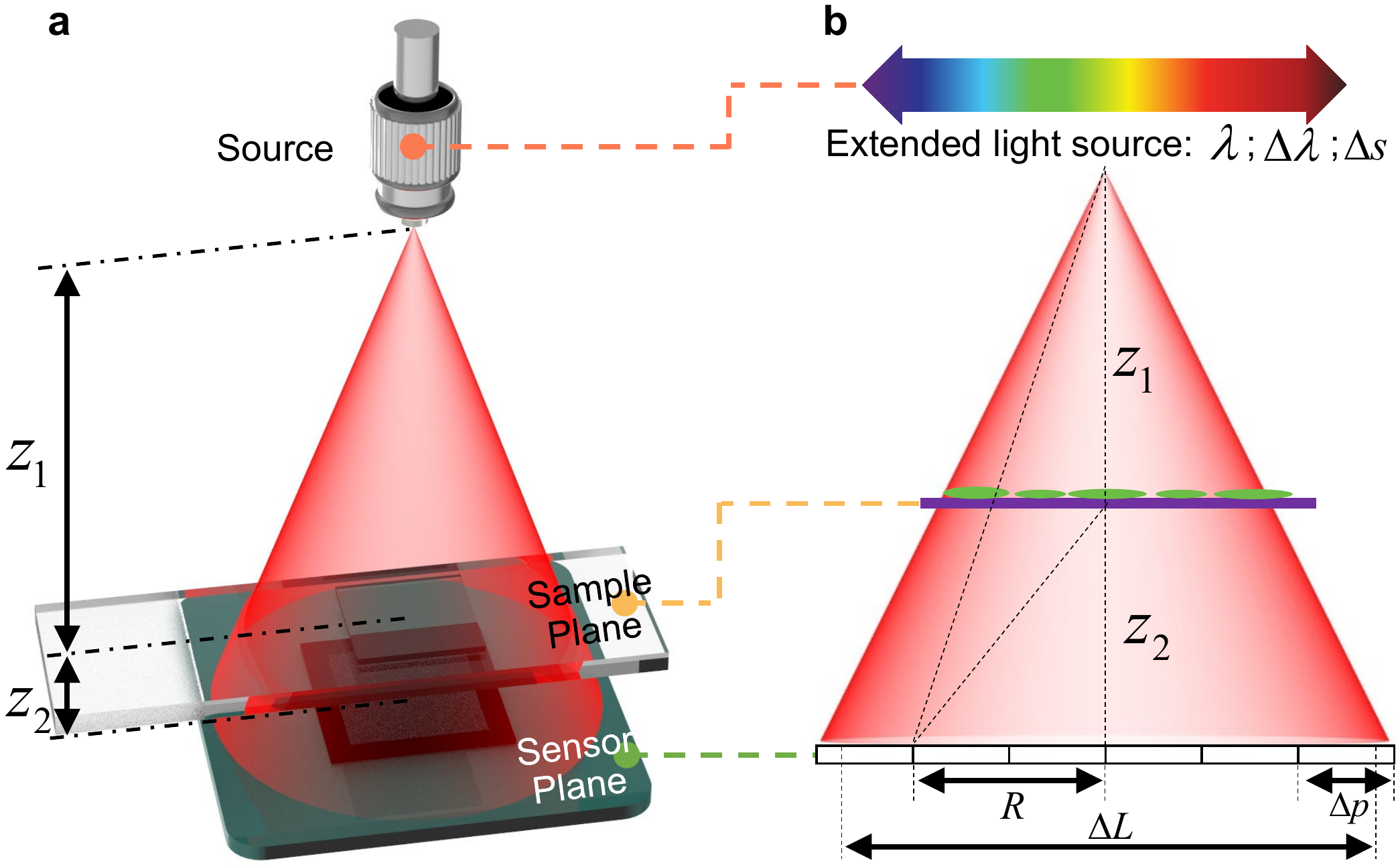}

\caption{Lens-free on-chip imaging. (a) General lens-free imaging experimental setup based on complementary metal-oxide semiconductor (CMOS) or charge-coupled device (CCD) image sensors. (b) Schematic of a lens-free holographic microscope. The sample is illuminated with wavelength $\lambda$, the spectral width $\Delta \lambda $, the diameter of the light-emitting area $\Delta s$. The diffraction pattern is registered by a sensor with pitch $\Delta p$ at a distance ${z_2}$.}
\label{fig:fig1}
\end{figure}

In the lens-free holographic microscope as depicted in Fig. \ref{fig:fig1}(a), the source can simply be a laser \cite{ozcan2008ultra,luo2016pixel,zhang2018lensfree}, a LED (an array of LEDs) \cite{tseng2010lensfree,kesavan2014high,ludwig2015calling,xiong2018optimized} or even a smartphone screen \cite{zheng2011epetri}. The coherent or partially coherent light illuminates the specimen, and then the scattered light and the transmitted light co-propagate in the same direction, finally forming interference fringes on the imaging device. In the ideal case, the sample should be placed on a sensor array which can directly capture the shadows of the objects and avoid the twin-image artifacts. However, due to the existence of protective glass covering the surface of the camera sensor, there is usually always a certain distance between the sample plane and the detector plane (typically ${\rm{0}}{\rm{.3}} - {\rm{2}}mm$) \cite{bishara2010lensfree,bishara2011holographic,ozcan2016lensless}. Since the distance is much larger than the wavelength, and the object information (including both amplitude and phase) is encoded into the diffraction patterns, which needs to be computationally reconstructed by phase retrieval and numerical back propagation algorithms.

As illustrated in the schematic diagram Fig. \ref{fig:fig1}(b) of the lens-free holographic microscope, neglecting the noise effect, the achievable resolution of LFOCDHM is determined by the maximum visualized radius $R$ of the diffraction patterns, which refer to the cut-off frequency of the transfer function. This transfer function can be further decomposed into five sub transfer functions, and the least cut-off frequency of the five transfer functions limits the maximum imaging resolution of LFOCDHM. The five sub transfer functions respectively correspond to the impact of the defocus distance, the limited temporal coherence length (the spectral width $\Delta \lambda $), the spatial coherence length (the diameter of light-emitting area $\Delta s$) of the source, the finite pixel size ($\Delta p$), and the finite extent of the image sub-FOV used for the reconstruction (the side length $\Delta L$). The absorption and phase transfer functions resulting from propagation are respectively denoted as $ATFP$ and $PTFP$. Then the temporal coherence transfer function, the spatial coherence transfer function, pixel size transfer function, the reconstructed region transfer function are severally expressed as $TCTF$, $SCTF$, $PSTF$, $RRTF$. Here, the latter four sub-transfer functions are mutually independent, and together have impacts on the final imaging results.

\subsection{Theoretical analysis of resolution in LFOCDHM}
\subsubsection{Influence of sample-to-sensor distance on imaging resolution}
In this subsection, we adopt the weak object approximation to simplify the mathematical formulation and linearize the phase retrieval problem \cite{kirkland2010advanced,zuo2017high}. The complex transmittance of a weak object can be represented as
\begin{equation}\label{weakobject}
\begin{gathered}
  t\left( {\bf{x}} \right) = a\left( {\bf{x}} \right){e^{i\phi \left( {\bf{x}} \right)}} \approx a\left( {\bf{x}} \right)\left[ {1 + i\phi \left( {\bf{x}} \right)} \right]\mathop \approx \limits^{a\left( {\bf{x}} \right) = {a_0} + \Delta a\left( {\bf{x}} \right)} {a_0} + \Delta a\left( {\bf{x}} \right) + i{a_0}\phi \left( {\bf{x}} \right),
\end{gathered}
\end{equation}
where $a\left( {\bf{x}} \right)$ is the absorption distribution with a mean value of ${a_0}$, $\phi \left( {\bf{x}} \right)$ is the phase distribution, ${\bf{x}} $ represents the two-dimensional coordinate (x,y) in spatial domain.
Taking Fourier transform of both sides of Eq. \ref{weakobject},  the Fourier spectrum of $t\left( {\bf{x}} \right)$ can be obtained as
\begin{equation}\label{FFTobject}
T\left( {\bf{u}} \right) = {a_0}\delta \left( {\bf{u}} \right) + A\left( {\bf{u}} \right) + i{a_0}\Phi \left( {\bf{u}} \right),
\end{equation}
where ${\bf{u}}$ is the two-dimensional coordinate in frequency domain, $\delta \left( {\bf{u}} \right)$ is the Dirac Delta function, $A\left( {\bf{u}} \right)$ and $\Phi \left( {\bf{u}} \right)$ respectively represent the Fourier spectrum of the absorption and phase distribution.

Before reaching the digital camera, the complex wave-front is propagated over the distance of ${z_2}$ in air (the medium of refractive index $ \approx {1}$) with the angular spectrum method \cite{goodman2005introduction}, which is equivalent to introducing an imaginary part into the transmitted complex wave-front in the Fourier domain:
\begin{equation}\label{FFTobjectpro}
\begin{aligned}
{W_{cam}}\left( {\bf{u}} \right) = {T}\left( {\bf{u}} \right)P\left( {\bf{u}} \right)= {a_0}\delta \left( {{\bf{u}}} \right)P\left( {{{\bf{u}}}} \right) + A\left( {{\bf{u}}} \right)P\left( {{{\bf{u}}}} \right) + i{a_0}\Phi \left( {{\bf{u}} } \right)P\left( {{{\bf{u}}}} \right),
\end{aligned}
\end{equation}
where $P\left( {\bf{u}} \right) = {e^{ik{z_{\rm{2}}}\sqrt {1 - {\lambda ^2}{{\left| {\bf{u}} \right|}^2}} }}$ represents the effect of defocus. At last, by calculating the convolution between ${W_{cam}}\left( {\bf{u}} \right)$ and its complex conjugate ${W'_{cam}}\left( {\bf{u}} \right) = {a_0}\delta \left( {{\bf{u}}} \right)P'\left( {  {{\bf{u}}}} \right) + A\left( {{\bf{u}}} \right)P'\left( {  {{\bf{u}}}} \right) - i{a_0}\Phi \left( {{\bf{u}}} \right)P'\left( {  {{\bf{u}}}} \right)$, we can get the intensity spectrum as:
\begin{equation}\label{FFTobjectInten}
\begin{aligned}
  {I_{cam}}\left( {\mathbf{u}} \right)& = {W_{cam}}\left( {\mathbf{u}} \right) \otimes {{W'}_{cam}}\left( {\mathbf{u}} \right) \hfill \\
  &\approx a_0^2P'\left( 0 \right)P\left( 0 \right)\delta \left( {\mathbf{u}} \right) + {a_0}A\left( {\mathbf{u}} \right)\left[ {P'\left( 0 \right)P\left( {\mathbf{u}} \right)}  +   {P\left( 0 \right)P'\left( {\mathbf{u}} \right)} \right]\hfill \\
    &+ ia_0^2\Phi \left( {\mathbf{u}} \right)\left[ {P'\left( 0 \right)P\left( {\mathbf{u}} \right) - P\left( 0 \right)P'\left( {\mathbf{u}} \right)} \right].
\end{aligned}
\end{equation}

\begin{figure}[!htb]
\centering
\includegraphics[width=\linewidth]{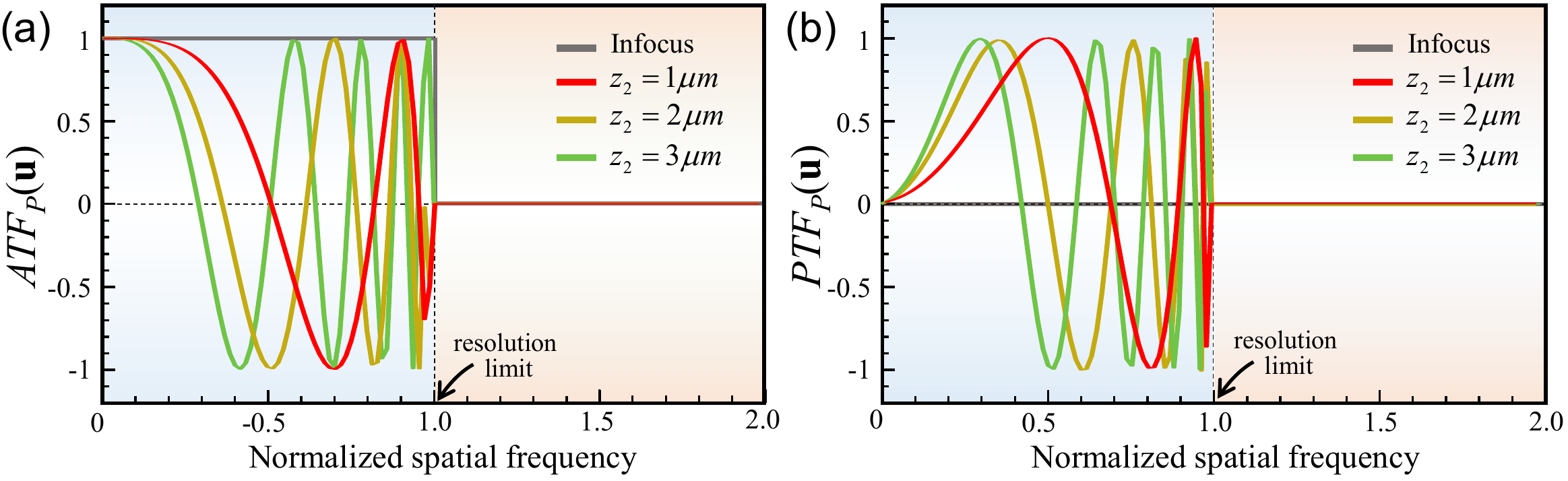}
\caption{The absorption transfer function $AT{F_p}\left( {\bf{u}} \right)$ (a) and phase transfer function $PT{F_p}\left( {\bf{u}} \right)$ (b) for various defocus distances. $\lambda {\rm{ = 600}}nm$, the spatial frequency coordinate is normalized against the resolution limit $1/\lambda$.}
\label{fig:fig2}
\end{figure}

\begin{figure}[!htb]
\centering
\includegraphics[width= \linewidth]{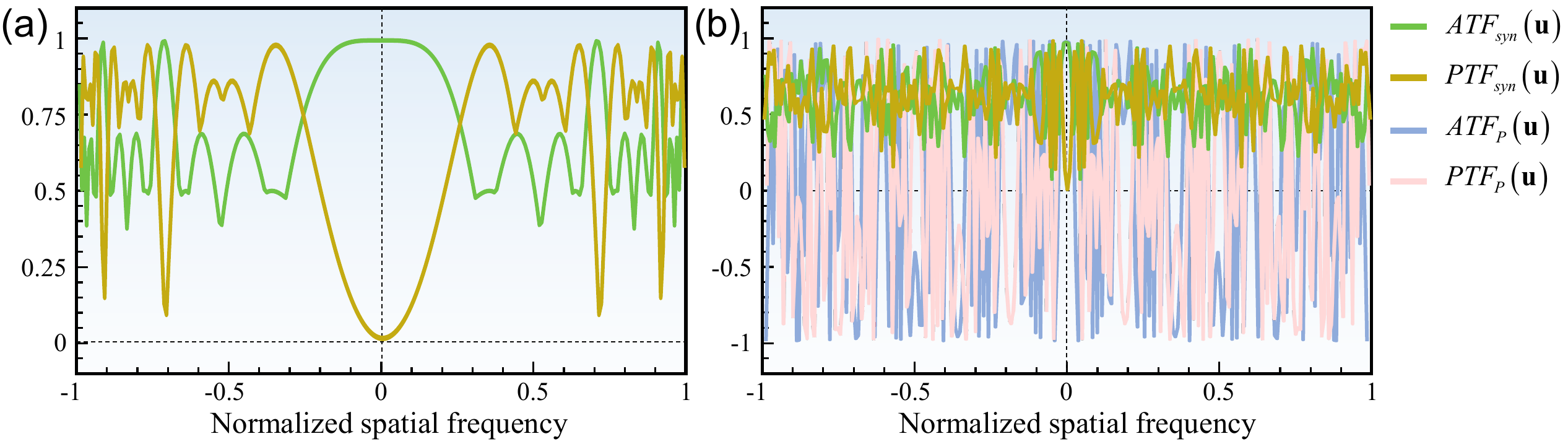}
\caption{(a) The synthesized absorption transfer function $AT{F_{syn}}\left( {\bf{u}} \right)$ and synthesized phase transfer function $PT{F_{syn}}\left( {\bf{u}} \right)$ with various defocus distances (${z_2} = 1,2,3\mu m$); (b) The absorption transfer function $AT{F_P}\left( {\bf{u}} \right)$ and phase transfer function $PT{F_P}\left( {\bf{u}} \right)$ with ${z_2} = 400 \mu m$; $AT{F_{syn}}\left( {\bf{u}} \right)$ and $PT{F_{syn}}\left( {\bf{u}} \right)$ with various defocus distances (${z_2} = 400,410,420\mu m$).}
\label{fig:fig3}
\end{figure}

In Eq. \ref{FFTobjectInten}, we neglect the high order convolution terms between $A\left( {\bf{u}} \right)$ and $\Phi \left( {\bf{u}} \right)$ to linearize the problem \cite{hamilton1984improved}. Thus, the absorption transfer function ($AT{F_p}$) and phase transfer function ($PT{F_p}$) of LFOCDHM with the defocus distance ${z_2}$ can be written as:
\begin{equation}\label{ATFPro}
\begin{aligned}
AT{F_p} = {a_0}\left[ {P'\left( 0 \right)P\left( {\mathbf{u}} \right) + P\left( 0 \right)P'\left( {\mathbf{u}} \right)} \right] = 2{a_0}\cos \left[ {k{z_2}\left( {1 - \sqrt {1 - {\lambda ^2}{{\left| {\mathbf{u}} \right|}^2}} } \right)} \right],\\
\end{aligned}
\end{equation}
\begin{equation}\label{PTFPro}
\begin{aligned}
PT{F_p} = a_0^2\left[ {P'\left( 0 \right)P\left( {\mathbf{u}} \right) - P\left( 0 \right)P'\left( {\mathbf{u}} \right)} \right] =  - 2a_0^2\sin \left[ {k{z_2}\left( {1 - \sqrt {1 - {\lambda ^2}{{\left| {\mathbf{u}} \right|}^2}} } \right)} \right]. \\
\end{aligned}
\end{equation}
The transfer functions of $AT{F_p}\left( {\bf{u}} \right)$ and $PT{F_p}\left( {\bf{u}} \right)$ with the wavelength $600 nm$ are shown in Fig. \ref{fig:fig2} for various defocus distances and the response value of them has been normalized to $0-1$. The sample-to-sensor distance ${z_2}$ varies from $1\mu m$ to $3\mu m$.
The simulation results of Fig. \ref{fig:fig2}(a) show that with the increase in defocus distance, the $AT{F_p}\left( {\bf{u}} \right)$ decreases earlier and the declining rate of these curves accelerates. Moreover, the increase in defocus distance also introduces higher oscillation frequency with more zero-crossings. The low responses of frequency around these zero-crossing points pose severe difficulties for the information reconstruction at these corresponding frequencies, suggesting that the information at these frequencies can no longer transfer into intensity and such high oscillation should be avoided as much as possible. Thus, for $AT{F_p}\left( {\bf{u}} \right)$, the smaller defocus distance will benefit for the reconstructed intensity image. However, for phase imaging $PT{F_p}\left( {\bf{u}} \right)$, Fig. \ref{fig:fig2}(b) shows that the response of frequency around the zero-point is always very low, suggesting the low-frequency phase can hardly transfer into intensity via defocusing. As the defocus distance getting large, the response at low frequencies gradually increases. In other words, large defocus distance is conducive to the recovery of the low-frequency phase information. Nevertheless, the accompanied high oscillation frequency will also introduce a large number of zero-crossing points. Thus, for the reconstruction of phase objects based on single sample-to-sensor distance, the selection of the defocus distance faces a fundamental tradeoff between low-frequency information reconstruction quality and the loss of frequency components. Thus, in general, multiple sample-to-sensor distances are required to construct a synthetic phase transfer function with high responses over a wider range of spatial frequencies:
\begin{equation}\label{TFpcom}
\begin{aligned}
\begin{array}{*{20}{c}}
  {AT{F_{syn}}\left( {\mathbf{u}} \right) = \frac{1}{{{N_{total}}}}\sum\limits_{i = 1}^{{N_{total}}} {\left| {AT{F_p}\left( {z_2^i,{\mathbf{u}}} \right)} \right|} } \\
  {PT{F_{syn}}\left( {\mathbf{u}} \right) = \frac{1}{{{N_{total}}}}\sum\limits_{i = 1}^{{N_{total}}} {\left| {PT{F_p}\left( {z_2^i,{\mathbf{u}}} \right)} \right|} }
\end{array},
\end{aligned}
\end{equation}
where $z_2^i$ represents the different defocus distances and $N_{total}$ is the total number of defocus planes. Under the same simulation conditions ($\lambda {\rm{ = 600}}nm$, $p {\rm{ = 300}}nm$, the spatial frequency coordinate is normalized against the resolution limit $1/ \lambda $.), the synthesized transfer functions of $AT{F_{syn}}\left( {\bf{u}} \right)$ and $PT{F_{syn}}\left( {\bf{u}} \right)$
are shown in Fig. \ref{fig:fig3}(a).

The simulation result of Fig. \ref{fig:fig3}(a) shows that the multi-height measurements can significantly reduce the number of zero-crossings by synthesization of transfer function. However, the recovery of the very low frequency (near zero frequency) phase component is still quite difficult. In the practical experiment, due to the cover glass of the sensor, the defocus distance usually exceeds $400\mu m$, and the oscillation frequency of the absorption transfer function $AT{F_P}\left( {\bf{u}} \right)$ and phase transfer function $PT{F_P}\left( {\bf{u}} \right)$ is extremely high, as shown in Fig. \ref{fig:fig3}(b). However, such a large distance can effectively reduce the low-response frequencies range, which is beneficial to recover the frequency components near zero-crossing points. Thus, when the defocusing distance reaches the order of several hundred microns, appropriately increasing $z_2$ can improve the reconstruction quality to some extent. Generally, when the components of the lens-free imaging system such as the light source and the sensor have been predetermined, multi-height measurements can optimize the synthetic transfer functions, which is beneficial for the intensity and phase reconstruction quality. But for single-height measurement, limited by the relatively large defocus distance, the influence of defocus distance on the reconstruction result can be neglected due to the rapid oscillation of the transfer functions. In the following part of this work, all simulations and experiments are carried out with single-height measurement to avoid the influence of multi-height selection on the reconstruction quality.

\subsubsection{Influence of temporal coherence of the illumination on imaging resolution}
In this section, we will analyze the influence of temporal coherence on the illumination on imaging resolution, which can be attributed to the temporal coherence transfer function ($TCTF$). Here, it is assumed that the temporal coherence is the only factor affecting the reconstruction resolution. Furthermore, in practical experiments, the ideal light source is difficult to obtain, and the LED light source is usually has a certain range of spectral width (for temporal coherence) and also luminous area (for spatial coherence). Supposing that the central wavelength $\lambda $, the spectrum width $\Delta \lambda $, the spectral distribution ${S_\lambda}\left( {{\lambda}_i} \right)$ are the predetermined parameters, and other system parameters are close to ideal values (do not affect the imaging resolution). If we further invoke the paraxial approximations \cite{zuo2017high}, the two transfer functions Eqs. (\ref{ATFPro}) and (\ref{PTFPro}) can be simplified as $AT{F_p} \approx 2{a_0}\cos \left( {\pi {z_2}\lambda {{\left| {\mathbf{u}} \right|}^2}} \right)$, $PT{F_p} \approx  - 2a_0^2\sin \left( {\pi {z_2}\lambda {{\left| {\mathbf{u}} \right|}^2}} \right)$. If the effect of spectral width of the illumination source is further taken into account, the absorption and phase transfer functions of LFOCDHM with the sample-to-sensor distance $z_2$ and the spectral width $\Delta \lambda$ can be can be represented as:
\begin{equation}\label{TFpt}
\begin{gathered}
\begin{array}{*{20}{c}}
{  AT{F_{p + t}}\left( {\mathbf{u}} \right) = 2{a_0}\int {{S_\lambda}\left( {\lambda  + {\lambda _x}} \right)} \cos \left[ {\pi {z_2}\left( {\lambda  + {\lambda _x}} \right){{\left| {\mathbf{u}} \right|}^2}} \right]d{\lambda _x} } \\
 { PT{F_{p + t}}\left( {\mathbf{u}} \right) =  - 2a_0^2\int {{S_\lambda}\left( {\lambda  + {\lambda _x}} \right)} \sin \left[ {\pi {z_2}\left( {\lambda  + {\lambda _x}} \right){{\left| {\mathbf{u}} \right|}^2}} \right]d{\lambda _x} }
\end{array},
\end{gathered}
\end{equation}
In most cases, the spectral distribution ${S_\lambda}$ can be approximated by an gaussian function:
\begin{equation}\label{SpectralCurve}
\begin{gathered}
{S_\lambda }\left( {{\lambda _i}} \right) = {e^{ - \frac{{{{\left( {{\lambda _i} - \lambda } \right)}^2}}}{{2{{\left( {\Delta \lambda {\text{/6}}} \right)}^{\text{2}}}}}}},
\end{gathered}
\end{equation}
where the mean value is $\lambda$ and the standard deviation is $\Delta \lambda/6$. Here standard deviation $\Delta \lambda/6$ is assumed to ensure that the normalized intensity of the wavelengths exceeding $\left[ {\lambda  - \Delta \lambda /2,\lambda  + \Delta \lambda /2} \right]$ will dip to $0.011$ and can be ignored. By incorporating the effect of temporal coherence, the transfer functions can be further expressed as the integrals over the full spectral range:
\begin{equation}\label{APTFpt}
\begin{gathered}
\begin{array}{*{20}{c}}
   { AT{F_{p + t}}\left( {\mathbf{u}} \right)=  2{a_0}\int_{ - \Delta \lambda /2}^{\Delta \lambda /2} {{e^{ - \frac{{\lambda _x^2}}{{2{{\left( {\Delta \lambda {\text{/6}}} \right)}^{\text{2}}}}}}}} \cos \left[ {\pi {z_2}\left( {\lambda  + {\lambda _x}} \right){{\left| {\mathbf{u}} \right|}^2}} \right]/\Delta \lambda d{\lambda _x}} \\
   {PT{F_{p + t}}\left( {\mathbf{u}} \right)=  - 2a_0^2\int_{ - \Delta \lambda /2}^{\Delta \lambda /2} {{e^{ - \frac{{\lambda _x^2}}{{2{{\left( {\Delta \lambda {\text{/6}}} \right)}^{\text{2}}}}}}}} \sin \left[ {\pi {z_2}\left( {\lambda  + {\lambda _x}} \right){{\left| {\mathbf{u}} \right|}^2}} \right]/\Delta \lambda d{\lambda _x}}
   \end{array}.
\end{gathered}
\end{equation}

We can find that Eq. \ref{APTFpt} is not integrable on real space, which will make this equation difficult to provide an analytical cut-off frequency expression. In addition, to give the theoretical cut-off frequency limit, in consideration of the ideal spectral distribution, we assume that ${S_\lambda}\left( {{\lambda}_i} \right)$ is a rectangular function, and then $AT{F_{p + t}}$ and $PT{F_{p + t}}$ will be noted as:
\begin{equation}\label{APTFpt2}
\begin{gathered}
\begin{aligned}
\begin{array}{*{20}{c}}
  AT{F_{p + t}}\left( {\mathbf{u}} \right) & = AT{F_P}\left( {\mathbf{u}} \right) sinc\left( {{z_2}\frac{{\Delta \lambda }}{2}{{\left| {\mathbf{u}} \right|}^2}} \right)\\
  PT{F_{p + t}}\left( {\mathbf{u}} \right)& = PT{F_P}\left( {\mathbf{u}} \right) sinc\left( {{z_2}\frac{{\Delta \lambda }}{2}{{\left| {\mathbf{u}} \right|}^2}} \right)
  \end{array}.
\end{aligned}
\end{gathered}
\end{equation}
\begin{figure}[!b]
\centering
\includegraphics[width=\linewidth]{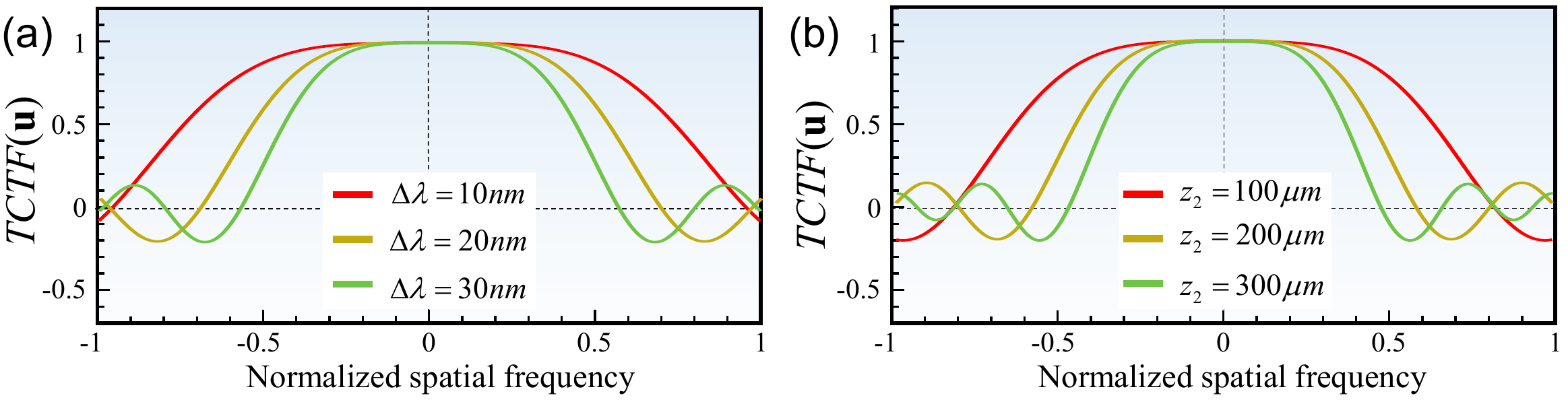}
\caption{The temporal coherence transfer function $TCTF\left( {\bf{u}} \right)$. (a) $TCTF\left( {\bf{u}} \right)$ for various spectral width $\Delta \lambda$ with the defocus distance ${z_2} = 200\mu m$. (b) $TCTF\left( {\bf{u}} \right)$ for various defocus distances with the spectral width $\Delta \lambda  = 30nm$.}
\label{fig:fig4}
\end{figure}

Based on Eq. \ref{APTFpt2}, the finite spectral width introduces an additional $sinc$ term to the transfer functions. Here, since the temporal coherence of light source play equally important role in the $AT{F_P}\left( {\bf{u}} \right)$ and $PT{F_P}\left( {\bf{u}} \right)$, we use $TCTF\left( {\mathbf{u}} \right)$ to represent the overall influence of finite spectral width:
\begin{equation}\label{TCTF}
TCTF\left( {\mathbf{u}} \right) = sinc\left( {{z_2}\frac{{\Delta \lambda }}{2}{{\left| {\mathbf{u}} \right|}^{\text{2}}}} \right).
\end{equation}

\begin{figure}[!b]
\centering
\includegraphics[width=0.8\linewidth]{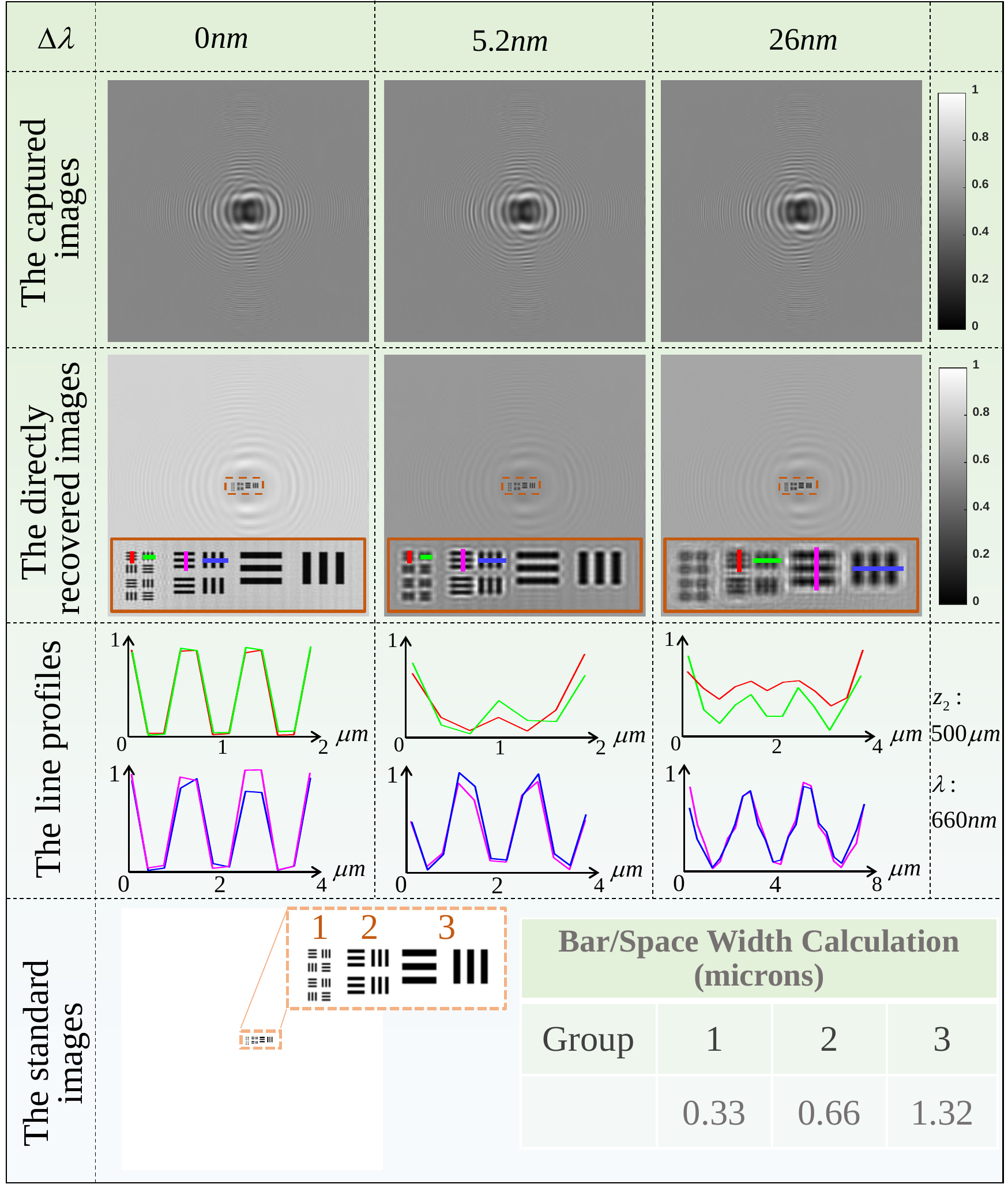}
\caption{The quantitative reconstruction results varying in the different spectral width $\Delta \lambda$. The simulation condition: ${z_2} = 500\mu m$, $\lambda  = 660nm$, $\Delta \lambda = 0, 5.2, 26 nm$. The first row: the raw images; The second row: the directly
reconstructed images with the angular spectrum method; The third row: the line profiles corresponding to the marks on the image in the second row; The forth row: the standard resolution target for the simulation.}
\label{fig:fig5}
\end{figure}

Then the temporal coherence transfer functions $TCTF\left( {\mathbf{u}} \right)$ for different spectral width $\Delta \lambda $ and various defocus distances are shown in Fig. \ref{fig:fig4}. In Fig. \ref{fig:fig4}(a), under the condition of ${z_2} = 200\mu m$, $\lambda  = 660nm$, the spectral width $\Delta \lambda $ varying from $10 nm$ to $30 nm$, as $\Delta \lambda $ gets wider, the frequency response decreases more rapidly and reaches zero earlier (at so-called the first zero-crossing or the first cut-off frequency). The response of the frequencies above the first cut-off frequency may slightly overshoot, but these frequency components are difficult to be recovered since the response is highly fluctuant. In contrast, for a given defocus distance ${z_2}$, higher temporal coherence (decreasing $\Delta \lambda $) provides a wider range of the high-response frequency regions and higher cut-off frequency, which is beneficial to improve the imaging resolution. In actual experiments, $\Delta \lambda $ usually is pre-defined parameter while the defocus distance ${z_2}$ is flexible, and thus the frequency response curves will be similar to those shown in Fig. \ref{fig:fig4}(b). The first cut-off frequency will gradually decrease as the defocus distance increases when the parameters of the light source are fixed. From Eq. \ref{TCTF}, we can deduce that the first cut-off frequency is at $\left| {\bf{u}} \right| = \sqrt {\frac{2}{{{z_2}\Delta \lambda }}}$, and the corresponding reconstructed half-pitch resolution is
\begin{equation}\label{ResolutionT}
q = \frac{1}{{2\left| {\mathbf{u}} \right|}} = \sqrt {\frac{{{z_2}\Delta \lambda }}{8}}.
\end{equation}

To verify the resolution limit resulting from the finite spectral width $\Delta \lambda $, we simulate a resolution target under conditions of ${z_2} = 500\mu m$, $\lambda  = 660nm$, as shown in Fig. \ref{fig:fig5}. From the line profiles in Fig. \ref{fig:fig5}, we can see that each element of the resolution target can be recovered when the light source is perfectly coherent, but the high-frequency elements gradually become blurred with the increase of $\Delta \lambda $. More specifically, when $\Delta \lambda $ is $5.2 nm$, the theoretical half-pitch resolution is $q = 0.57 \mu m$, which coincides well with the simulation result shown in Fig. \ref{fig:fig5}. For $\Delta \lambda = 26 nm$, the elements of group 3 can be distinguished easily, but elements of group 2 are barely discernable. According to Eq. \ref{ResolutionT} (the theoretical resolution $q = 1.27 \mu m$), group 2 of the target should be completely indistinguishable, so the slightly discernible elements may result from the non-zero responses of the transfer function beyond the first cut-off frequency, as shown in Fig. \ref{fig:fig4}.

In summary, the temporal coherence of illumination have an impact on the ultimate imaging resolution of the LFOCDHM system. Increasing temporal coherence of the source by using a laser, or insert a narrow band-pass filter in front of the source can directly reduce its influence on the resolution. When the light source of the system is determined ($\Delta \lambda$ is a constant value), it should be guaranteed that the object-to-sample distance $z_2$ must be smaller than ${2}{\lambda ^{2}}{/}\Delta \lambda$ (guarantee $q$ is smaller than $\lambda /2$) so that the temporal coherence of the source does not influence the final resolution, and the reconstructed resolution will be only affected by the ideal coherent diffraction limit ($\lambda /2$). For example, when the spectrum width of illumination source is about $20nm$ and the ideal half-pitch resolution limit is $0.5 \mu m$, the object-to-sample distance $z_2$ should be smaller than $100 \mu m$ ideally. However, for imaging phase objects, $z_2$ should not be too small to guarantee sufficient responses of the phase transfer function, which is crucial to the recovery accuracy of low-frequency phase information. As mentioned earlier, due to the manufacturing technology of sensors, the defocusing distance $z_2$ cannot go below 300 $\mu$m. When the distance $z_2$ cannot be small enough, we should use a light source with higher temporal coherence (narrower spectral width $\Delta \lambda$) to guarantee the diffraction-limited imaging resolution.

\subsubsection{Influence of spatial coherence of the illumination on imaging resolution}
In this section, we will analyze the influence of spatial coherence on the illumination on imaging resolution, which can be attributed to the spatial coherence transfer function ($SCTF$). In addition to the temporal coherence of the light source, the spatial coherence also affects the reconstructed resolution. Same as before, assuming that the reconstructed resolution is only affected by the spatial coherence of the light source. We also assume that the sample is illuminated by the light emitting from a spatially incoherent delta-correlated light source (any two different points in the source plane are uncorrelated), and the acquired hologram can be interpreted as an incoherent superposition of all partial holograms arising from all light source points. In other words, the influence of the spatial coherence can be modeled as a convolution of the ideal in-line hologram ${I}\left( \bf{x} \right)$ (arising from the central point source) with a properly resized source intensity distribution ${S_s}\left( {{{\mathbf{x}}_s}} \right)$ \cite{feng_resolution_2017}.
\begin{equation}\label{IcapSum}
\begin{gathered}
  {I_{cap}}\left( \bf{x} \right) = I\left( \bf{x} \right) \otimes \left[{\left( {\frac{{{z_1}}}{{{z_2}}}} \right)^2} \times {S_s}\left( {\frac{{{z_1}}}{{{z_2}}}\bf{x}} \right)\right] = I\left( \bf{x} \right) \otimes PSF\left( \bf{x} \right),\hfill \\
\end{gathered}
\end{equation}
where $\bf{x}$ represents the coordinates in the imaging sensor plane, ${{{\mathbf{x}}_s}}$ are the coordinates in the illumination plane. Without loss of generality, the scaled factor ${\left( {{z_1}{/}{z_2}} \right)^2}$ can be neglected. According to Eq. \ref{IcapSum}, assuming that the illumination source is circular with a diameter of $\Delta s$, the spatial coherence transfer function ($SCTF$) can be expressed as:
\begin{equation}\label{APTFpts}
\begin{gathered}
SCTF\left( {\mathbf{u}} \right) = \mathcal{F}\left( {PSF} \right) = \frac{{\sin \left( {\pi \frac{{{z_2}{\Delta s}}}{{{z_1}}}{\left| {\mathbf{u}} \right|}} \right)}}{{\pi \frac{{{z_2}{\Delta s}}}{{{z_1}}}{\left| {\mathbf{u}} \right|}}} = sinc \left( {\frac{{{z_2}{\Delta s}}}{{{z_1}}}{\left| {\mathbf{u}} \right|}} \right).
\end{gathered}
\end{equation}

\begin{figure}[!htb]
\centering
\includegraphics[width=\linewidth]{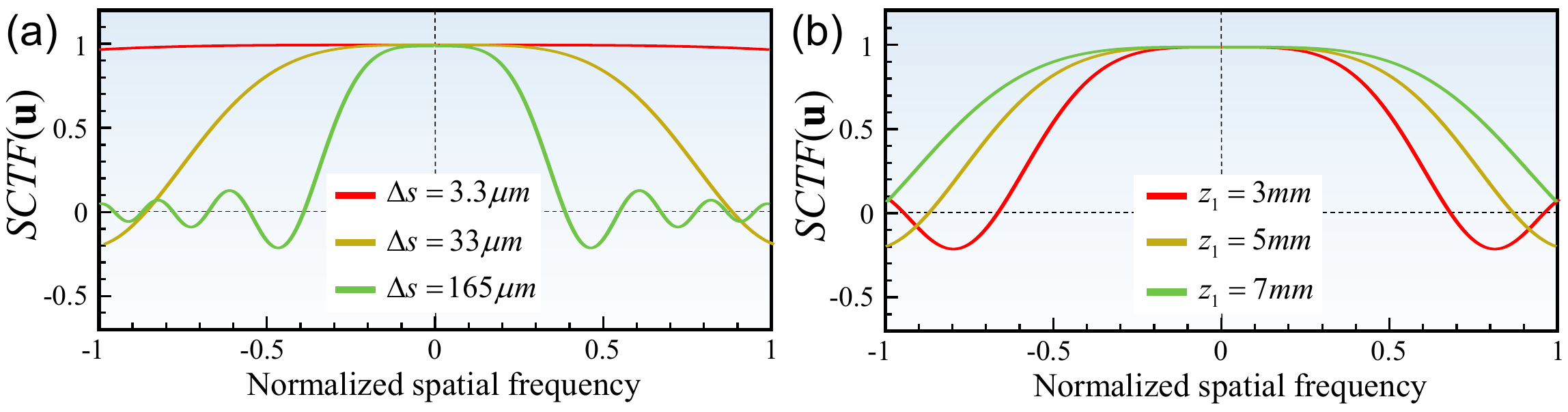}
\caption{The spatial coherence transfer function $SCTF\left( {\bf{u}} \right)$. (a) $SCTF\left( {\bf{u}} \right)$ for different illumination areas. (${z_1} = 5 mm$, ${z_2} = 200\mu m$) (b) $SCTF\left( {\bf{u}} \right)$ for various defocus distances with the diameter of the light-emitting zone $\Delta s = 33\mu m$.(${z_2} = 200\mu m$)}
\label{fig:fig6}
\end{figure}

The simulation results of the transfer function $SCTF\left( {\mathbf{u}} \right)$ for different source sizes and defocus distances are shown in Fig. \ref{fig:fig6}(a) and \ref{fig:fig6}(b). In Fig. \ref{fig:fig6}(a), $\Delta \lambda  \to 0$, $\lambda  = 660nm$, ${z_1} = 5 mm$, ${z_2} = 200\mu m$, $\Delta s = 3.3,33,165\mu m$ are given to analyze the resolution limit resulting from the spatial coherence. From the simulation results of Fig. \ref{fig:fig6}(a), the effect of the spatial coherence on the reconstruction resolution will reduce as the illumination area getting smaller. From the curves of $SCTF\left( {\bf{u}} \right)$ in Fig. \ref{fig:fig6}(a), while $\Delta s$ gets larger, the response of the transfer function decrease earlier and reach the first cut-off frequency more rapidly.

In actual experiments, when the illumination source is determined, the diameter of the luminous area ($\Delta s$) is unalterable. Under such condition, in order to improve the spatial coherence, we can increase the shrink ratio of ${{z_1}/{z_2}}$ to reduce the effective illumination area, alternatively. In our simulations, the system parameters are $\Delta \lambda  \to 0$, $\lambda  = 660nm$, ${z_1} = 3, 5, 7 mm$, ${z_2} = 200\mu m$, $\Delta s = 33\mu m$, and the frequency response curves are shown in Fig. \ref{fig:fig6}(b). From these curves, we can observe that larger ${{z_1}/{z_2}}$ will increase the first cut-off frequency, and thus,  improve the reconstruction resolution. Based on Eq. \ref{APTFpts}, we can derive that the first cut-off frequency is $\left| {\bf{u}} \right|  = \frac{{{z_1}}}{{{z_2}{\Delta s}}}$, and the corresponding reconstructed half-pitch resolution is
\begin{equation}\label{ResolutionS}
q = \frac{1}{{2\left| {\mathbf{u}} \right|}} = \frac{{{z_2}{\Delta s}}}{{2{z_1}}}.
\end{equation}

\begin{figure}[!htb]
\centering
\includegraphics[width=0.8\linewidth]{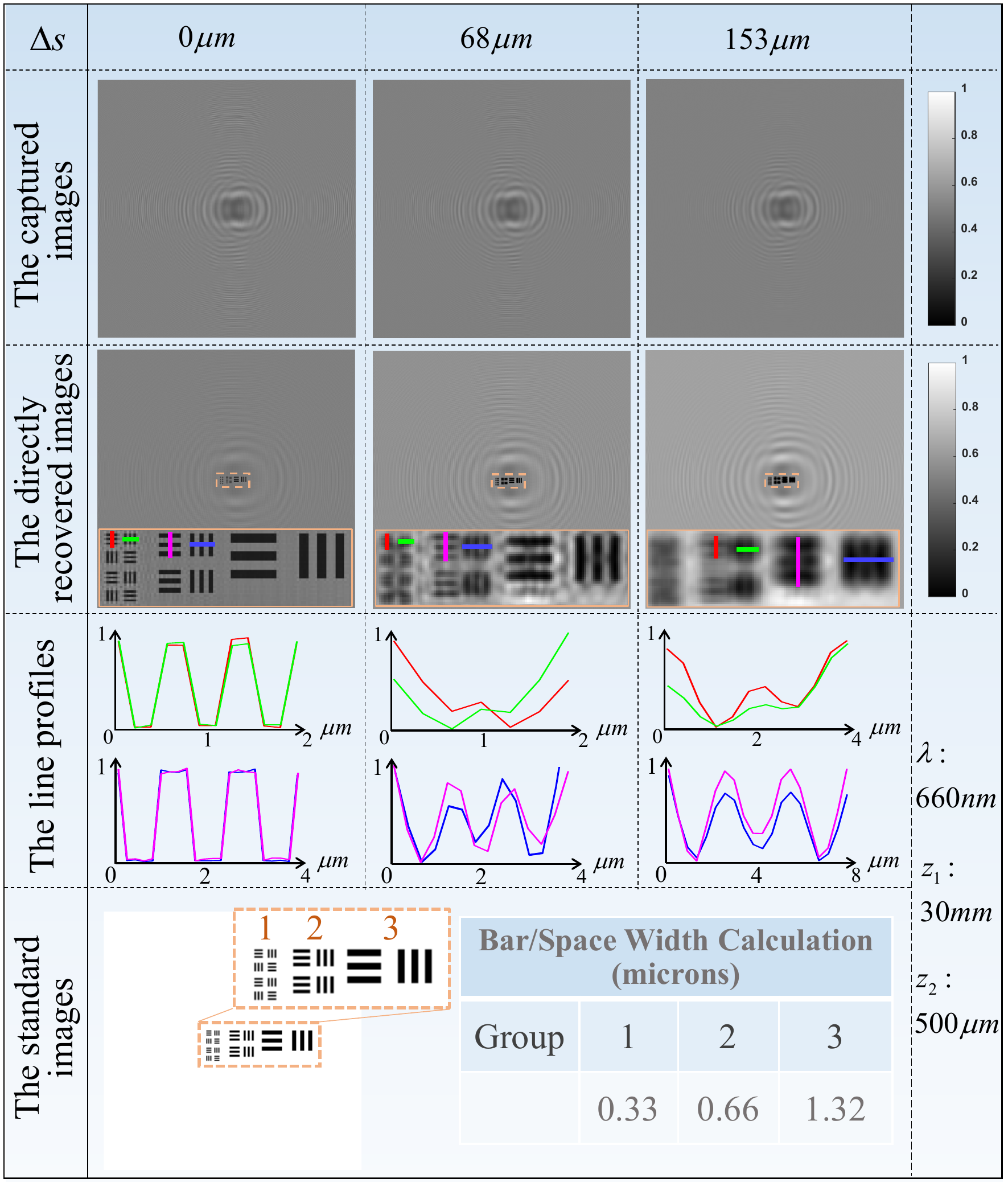}
\caption{The quantitative reconstruction results varying in the diameters of illumination source $\Delta s$. The simulation condition: ${z_2} = 30m m$, ${z_2} = 500\mu m$, $\lambda  = 660nm$, $\Delta s = 0, 68, 153 \mu m$. The first row: the raw images; The second row: the directly
reconstructed images with the angular spectrum method; The third row: the line profiles corresponding to the marks on the image in the second row; The forth row: the standard resolution target for the simulation.}
\label{fig:fig7}
\end{figure}

This reconstruction resolution involves many parameters and factors according to Eq. \ref{ResolutionS}. In Fig. \ref{fig:fig7}, ${z_1} = 30 mm$, ${z_2} = 500\mu m$ are given to verify the resolution limit. In Fig. \ref{fig:fig7}, when ${\Delta s}$ gradually increases, the reconstruction resolution will get worse correspondingly. For example, when ${\Delta s} = 68 \mu m$, the theoretical resolution is $0.57 \mu m$, and the corresponding simulation result is $0.66 \mu m$ which is lower than that of the ideal illumination ${\Delta s} \to 0$. If ${\Delta s}$ further increases to $153 \mu m$, the resolution reduced to $1.32 \mu m$, which agrees with the theoretical value $1.28 \mu m$.

From the above analysis, we know that the spatial coherence may affect the ultimate imaging resolution of the LFOCDHM system, which is associated with the ratio $z_2 / z_1$ and ${\Delta s}$.
Thus, in the lens-free experimental setups, when the LED is used as a light source, there are several ways to  improve the spatial coherence and reduce its effect on imaging resolution. On the one hand, we can insert a small pin-hole in front of the source to reduce the source size. On the other hand, we can reduce to ratio $z_2 / z_1$ to reduce the effective size of the source. As we mentioned earlier, the object-to-sample distance $z_2$ cannot be too small, so we can the source-to-sample distance $z_1$ instead. All these experimental manipulations are to avoid the effect of the poor spatial coherence on the reconstruction resolution, and guarantee the diffraction-limited imaging resolution [$q$ (Eq. \ref{ResolutionS}] is smaller than $\lambda /2$). For example, when the diameter ${\Delta s}$ of illumination source is about $200 \mu m$ and the ideal half-pitch resolution limit is $0.5 \mu m$, ratio $z_2 / z_1$ must be smaller than $1/200$ theoretically.  However, for actual imaging objects, $z_2$ is usually larger than $400 \mu m$, and thus, to guarantee sufficient responses of the transfer function, $z_1$ must be larger than $80 mm$. Consequently, for an established lens-free microscopic imaging system, the effect of spatial coherence can be avoided as far as possible by increasing $z_1$.

\subsubsection{Influence of sensor pixel size on imaging resolution}

In lens-free imaging system, the pixel size is a key factor influencing the achievable spatial resolution. Assuming that the actual pixel size and resolution of the camera respectively are $\Delta p$ and $m \times n$, the finest feature to be reconstructed corresponds to the half-pitch resolution $\Delta p /w$, which is $w \left( {w \geqslant 1} \right)$ times smaller than the actual sampling rate of the camera. The number of pixels of the reconstructed image is $M \times N$. The ideal pixel aliasing can be interpreted as a procedure that the ideal image is first pixel binning and then sub-sampled. Specifically, the pixel binning effect can be modeled as:
\begin{equation}\label{aliasing}
{I_{bin}}\left( {\mathbf{x}} \right) = {I_{bin}}\left( {x,y} \right) = \frac{1}{{{w^2}}}\sum\limits_{{w_y} = 0}^{w - 1} {\sum\limits_{{w_x} = 0}^{w - 1} {I\left( {x - {w_x},y - {w_y}} \right)} },
\end{equation}
where $I\left( \bf{x} \right)$ is the ideal image, $\bf{x}$ is two-dimensional coordinates on camera plane. Thus, in the frequency domain, this process can be represented as:
\begin{equation}\label{aliasing}
\begin{gathered}
\begin{aligned}
O_{bin}\left( {\mathbf{u}} \right) &= \mathcal{F}\left( {I_{bin}\left( \bf{x} \right)} \right) = \mathcal{F}\left( {I\left( \bf{x} \right)} \right)PSTF\left( {\mathbf{u}} \right) \\
&= O\left( {\mathbf{u}} \right)PSTF\left( {\mathbf{u}} \right),
\end{aligned}
\end{gathered}
\end{equation}
where $O_{bin}\left( {\mathbf{u}} \right)$ and $O\left( {\mathbf{u}} \right)$ is the Fourier transform of $I_{bin}\left( \bf{x} \right)$ and $I\left( \bf{x} \right)$, respectively. $PSTF\left( {\mathbf{u}} \right)$ is the transfer function corresponding to the pixel binning, which takes the following form:
\begin{equation}\label{PSTF}
PSTF\left( {\mathbf{u}} \right) = PSTF\left( {{u_x},{u_y}} \right) = \frac{1}{{{w^2}}}\sum\limits_{\alpha = 1}^w {\sum\limits_{\beta = 1}^w {\exp \left\{ {j\pi \left[ {(w - 1){u_x} + (w - 1){u_y}} \right]} \right\}} }.
\end{equation}

When ${u_x} =  \pm \frac{{{r_x}}}{w}$ or ${u_y} =  \pm \frac{{{r_y}}}{w}$ or $w=1$ (${r_x}$, ${r_y}$ is a positive integer not greater than $w/2$ and the frequency has been normalized to $- 1/2 \sim 1/2$.), $PSTF$ will be zero, suggesting that the corresponding spectral information is lost. Thus, the normalized first cut-off frequency will be $1/w$. Due to the previous assumptions that the ideal theoretical half-pitch resolution is $\Delta p/{w}$, the resolution limit after aliasing can be noted as:
\begin{equation}\label{ResolutionP}
q = \Delta p.
\end{equation}

For the second step, the sampling process is that the ideal images are sampled at uniform intervals ($w$ pixels). One way
to model sampling is to multiply $I\left( \bf {x} \right)$ by a sampling function $S_w\left( \bf {x} \right)$ equal to a train
of impulses $w$ units apart \cite{gonzalez1977digital}. That is
 \begin{equation}\label{SamplingSpatial}
{I_{Sam}}\left( \bf {x} \right) = I_{ali}\left( \bf {x} \right) \cdot {S_w}\left( \bf {x} \right),
\end{equation}
where ${I_{Sam}}\left(  \bf {x} \right)$ is the image after sampling, ${S_w}\left( \bf {x} \right)$ is the two-dimensional sampling function. Here ${S_w}\left( {\mathbf{x}} \right) = {S_w}\left( {x,y} \right) = \sum\limits_{\alpha =  - M/2}^{M/2 - 1} {\sum\limits_{\beta =  - N/2}^{N/2} {\delta \left( {x - \alpha w,y - \beta w} \right)} } $. In the Fourier space, Eq. \ref{SamplingSpatial} can be written as:
\begin{equation}\label{SamplingFourier}
{O_{Sam}}\left( {\mathbf{u}} \right) = O_{ali}\left( {\mathbf{u}} \right) \otimes {\tilde S_w}\left( {\mathbf{u}} \right),
\end{equation}
where ${\tilde S_w}\left( {\mathbf{u}}  \right) = \sum\limits_{\alpha = 0}^{w - 1} {\sum\limits_{\beta = 0}^{w - 1} {\delta \left( {{u_x} - {\alpha} \frac{M}{w},{u_y} - {\beta}\frac{N}{w}} \right)} } $. In discrete numerical calculation, the dimension of the captured image is different from that of the original image, so the sampling process can be written in the form of matrix: ${O_{cap}} = {M_{left}}{O_{ali}}{M_{right}}$, where ${M_{left}}$ is a $m \times M$ matrix, and ${M_{right}}$ is a $n \times N$ matrix. Concretely, ${M_{left}} = \left[ {\overbrace {{\mathbf{A}} \cdots {\mathbf{A}}}^w} \right]$,
${\mathbf{A}} = \left[ {\begin{array}{*{20}{c}}
  {{{\mathbf{A}}_{\mathbf{1}}}}&{{{\mathbf{A}}_2}} \\
  {{{\mathbf{A}}_2}}&{{{\mathbf{A}}_{\mathbf{1}}}}
\end{array}} \right]$. When ${{\mathbf{I}}_A}$ is the $\frac{M}{{{\text{2}}w}} \times \frac{M}{{{\text{2}}w}}$ unit matrix, then ${{\mathbf{A}}_1}$ and ${{\mathbf{A}}_2}$ can be denoted by ${{\mathbf{A}}_1} = \frac{{{{\mathbf{{\rm I}}}_A} - {{\left( { - 1} \right)}^w}{{\mathbf{I}}_A}}}{2}$, ${{\mathbf{A}}_2}{\text{ = }}\frac{{{{\mathbf{{\rm I}}}_A}{\text{ + }}{{\left( { - 1} \right)}^w}{{\mathbf{I}}_A}}}{2}$. Analogously, ${M_{right}} = \left[ {w\left\{ {\begin{array}{*{20}{c}}
  {\mathbf{B}} \\
   \vdots  \\
  {\mathbf{B}}
\end{array}} \right.} \right]$,
${\mathbf{B}} = \left[ {\begin{array}{*{20}{c}}
  {{{\mathbf{B}}_{\mathbf{1}}}}&{{{\mathbf{B}}_2}} \\
  {{{\mathbf{B}}_2}}&{{{\mathbf{B}}_{\mathbf{1}}}}
\end{array}} \right]$, ${{\mathbf{B}}_1} = \frac{{{{\mathbf{{\rm I}}}_B} - {{\left( { - 1} \right)}^w}{{\mathbf{I}}_B}}}{2}$, ${{\mathbf{B}}_2} = \frac{{{{\mathbf{{\rm I}}}_B} + {{\left( { - 1} \right)}^w}{{\mathbf{I}}_B}}}{2}$, where ${{\mathbf{I}}_B}$ is a $\frac{N}{{{\text{2}}w}} \times \frac{N}{{{\text{2}}w}}$ unit matrix. The process shows that the high-frequency information will be mixed into the low-frequency domain.  

\begin{figure}[!htb]
\centering
\includegraphics[width=0.8\linewidth]{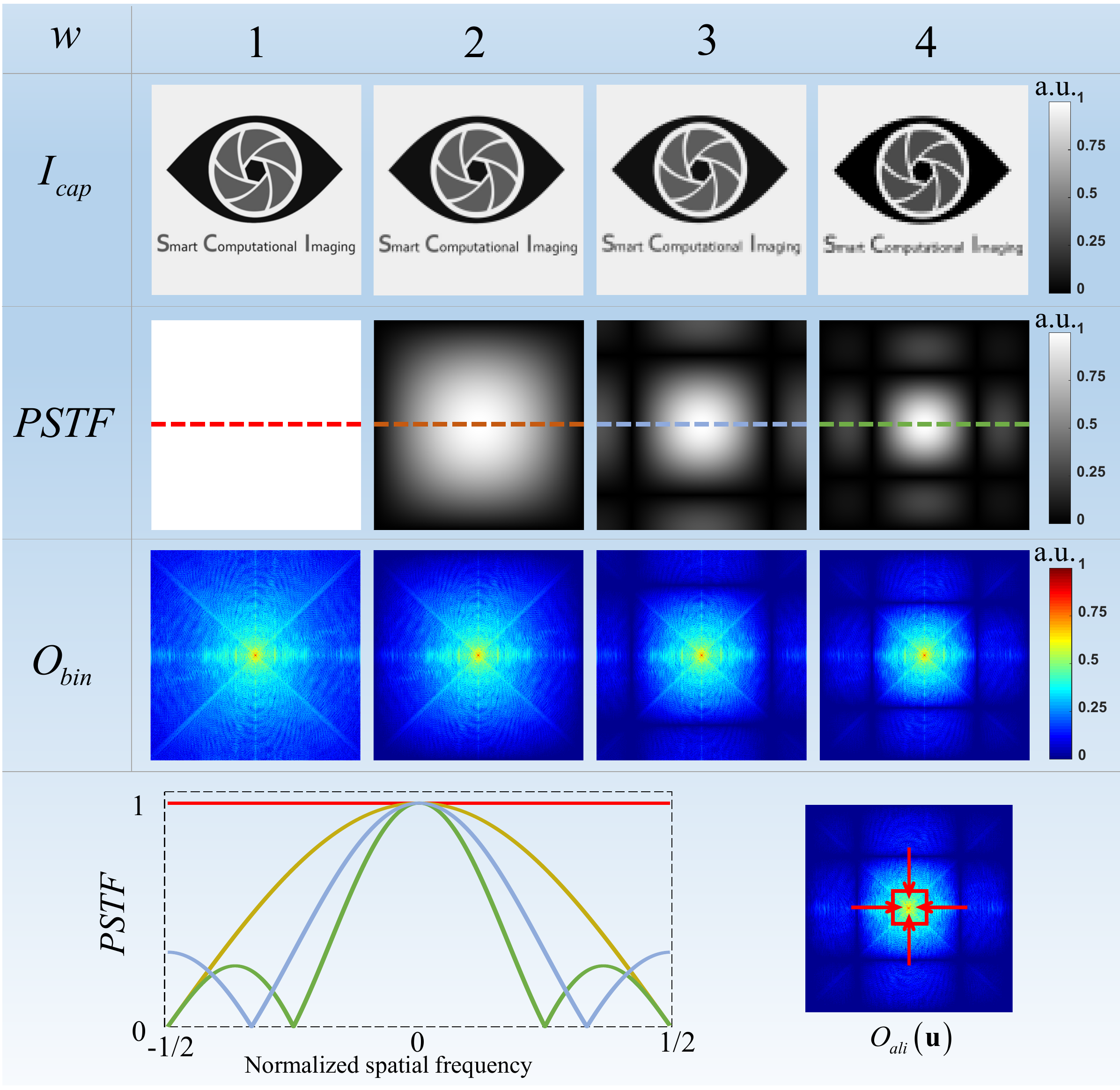}
\caption{The simulation results with the down-sampling factors $w = 1,2,3,4$. The first row: the raw captured images; The second row: the corresponding pixel aliasing transfer functions; The third row: the Fourier spectrum of images with aliasing; The forth row: Left: the line profiles of the pixel aliasing transfer functions, Right: the sampling process with $w=4$.}
\label{fig:fig8}
\end{figure}

To show the information aliasing and spectrum loss resulting from the finite pixel size, the simulation results with the down-sampling factors $w = 1,2,3,4$ are illustrated in Fig. \ref{fig:fig8}. On the other hand, $w$ can also be regarded as the resolution up-sampling factor for the pixel SR reconstruction algorithm from LR intensity measurements. The line curves of $PSTF\left( {\mathbf{u}} \right)$ show that when $w$ gradually increases, the more criss-crossed frequency gaps will appear, suggesting the information around these frequencies will exceptionally difficult to be recovered. When $w = 2$, $PSTF\left( {\mathbf{u}} \right)$ tends to zero only at the highest frequency (the periphery of the Fourier spectrum). When $w > 2$, more spectral information at interlaced regions in $PSTF\left( {\mathbf{u}} \right)$ becomes zero. The lower right of Fig. \ref{fig:fig8} shows the Fourier spectrum $O_{bin}\left( {\mathbf{u}} \right)$ after pixel binning with $w = 4$, and the red rectangular area ($\frac{M}{w} \times \frac{N}{w}$) has the same dimensional size with the captured image. The whole process shows that the high-frequency information will be mixed into the low-frequency domain within the red rectangle, and the aliasing problem will be more serious when $w$ getting larger. For normal pixel size of the current image sensor (typically $0.8 - 5 \mu m$), the pixel aliasing is a key limiting factor directly affect the imaging resolution of the LFOCDHM system. When the resolution of the object to be reconstructed (by pixel SR algorithms \cite{zheng2010sub,bishara2010lensfree,bishara2011holographic,luo2016propagation}) is $w$ times higher than that limited by the original pixel size, the number of the captured raw LR images (theoretical amount of information) will linearly increase with a factor of $w^{2}$ \cite{miao1998phase}.

\subsubsection{Influence of the finite extent of reconstructed sub-FOV on imaging resolution}

\begin{figure}[!b]
\centering
\includegraphics[width=0.8\linewidth]{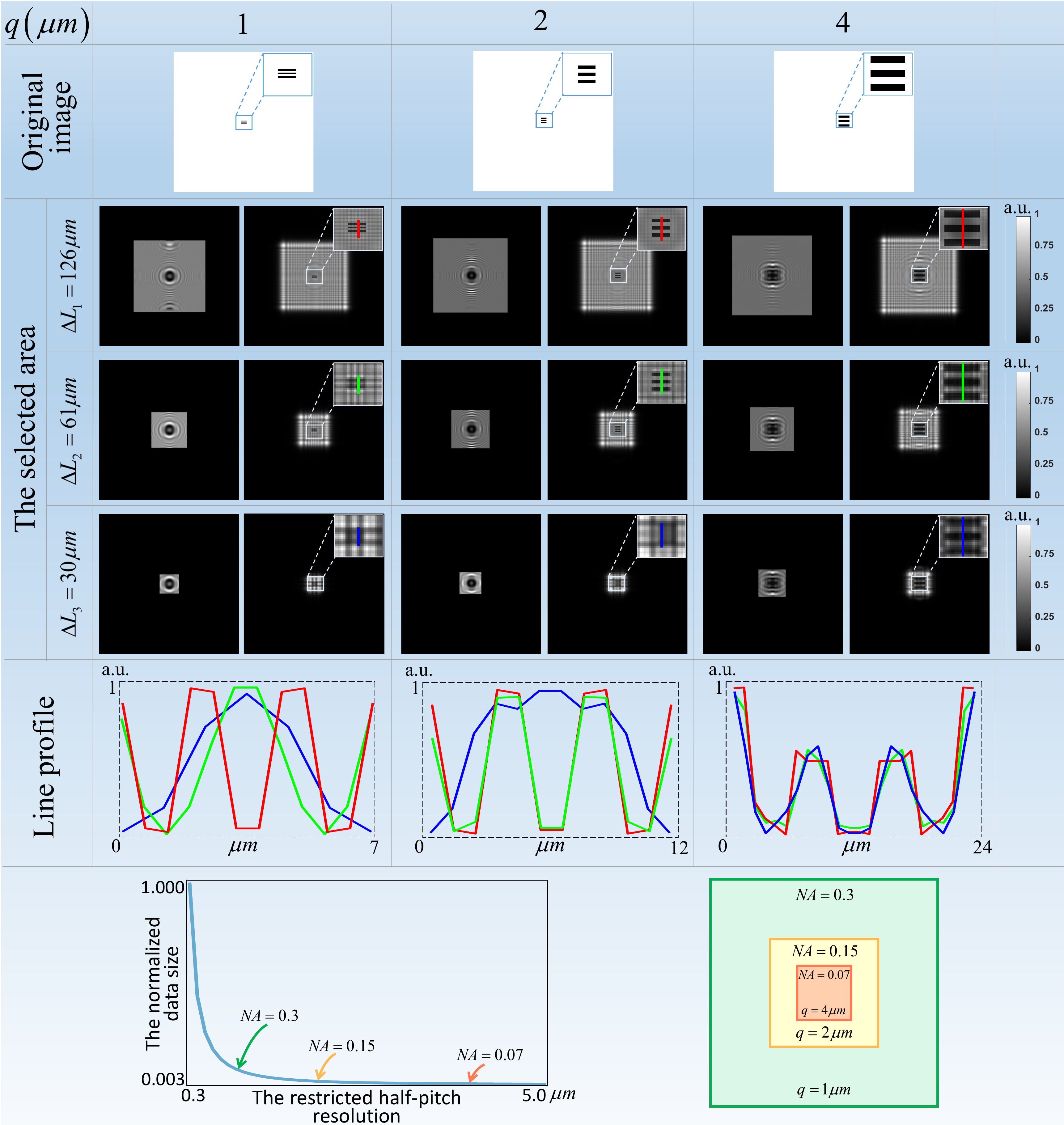}
\caption{ From the first to fifth row: The simulation results with different reconstructed area sizes ($\Delta {L_1}=126 \mu m$, $\Delta {L_2}=61 \mu m$, $\Delta {L_3}=30 \mu m$). The last row: Left: The half-pitch-resolution-dependent curve of the reconstructed area size; Right: The relative size of reconstructed region corresponding to different half-pitch resolution.}
\label{fig:fig9}
\end{figure}

As we mentioned in the introduction, one of the most important advantages of the LFOCDHM is the large effective numerical aperture $ \sim 1$ over a very large FOV because the sample-to-sensor sensor distance is much smaller than the size of the imaging sensor. However, in practice, due to the limited processing capability and memory of the computer, usually each raw image is divided into several subregions for the holographic reconstruction, and the reconstructed sub-images are then stitched together to obtain the whole-FOV image. Due to the limited extent of the selected reconstructed area (assuming that the side length of the sub-FOV is $\Delta L$), some high-angle diffraction patterns corresponding to the high-frequency of the object will not be included in the reconstructed area, leading to the reduction of imaging resolution. We attribute the effect of finite extent of reconstructed sub-FOV on the Fourier spectrum to the transfer function $RRTF$, and the cut-off frequency of $RRTF$ is $\left| {\mathbf{u}} \right| = \frac{{{\Delta L}/2}}{{\lambda \sqrt {{z_2}^2 + {{\left( {{\Delta L}/2} \right)}^2}} }}$. Thus, the reconstructed half-pitch resolution is determined by the effective NA of the LFOCDHM system, which can be represented as the ratio between ${\Delta L}/2$ and $\sqrt {{z_2}^2 + {\left( {\Delta L}/2 \right)}^2}$ (as shown in Fig. \ref{fig:fig1}), and the restricted half-pitch resolution is
\begin{equation}\label{ResolutionROI}
q = \frac{1}{{2\left| {\mathbf{u}} \right|}} = \frac{{\lambda \sqrt {4z_2^2 + {{\Delta L}^2}} }}{{2{\Delta L}}}.
\end{equation}
According to Eq. \ref{ResolutionROI}, in order to achieve the half-pitch resolution $q$, the side length of reconstructed sub-FOV should meet the following requirement:
\begin{equation}\label{ROIL}
\Delta L \geqslant \frac{{2{z_2}\lambda }}{{\sqrt {4{q^2} - {\lambda ^2}} }}.
\end{equation}

In the simulation, we use $\lambda  = 600nm$, ${z_2} = 200\mu m$, $\Delta p = 1\mu m$, and the theoretical half-pitch resolution ${q} = 1,2,4\mu m$ can be calculated to verify the influence of the reconstructed area on the resolution. In Fig. \ref{fig:fig9}, we can find that when the side length is ${\Delta L_1} = 126\mu m$, the maximum half-pitch resolution is about $1\mu m$. However, when $\Delta L$ is getting smaller, the maximum half-pitch resolution will gradually decrease, e.g., when the side length is ${\Delta L_2} = 61\mu m$, the half-pitch resolution will reduce to $2\mu m$. As shown in Fig. \ref{fig:fig9}, the reconstructed area size almost increases exponentially with the improvement of the half-pitch resolution. Thus, for example, when the sample-to-sensor distance is $400 \mu m$, in order to achieve the high imaging resolution close to the diffraction limit (e.g. NA $ \sim 0.8$), the slide length of the reconstructed sub-FOV should be at least $2845 \mu m$, which again brings a big challenge to the computational efficiency and memory requirement (especially when the pixel SR algorithm is used).

Furthermore, for each reconstruction of sub-FOV, only very limited central region can achieve the expected resolution. For the rest part, the region more close to the border will have lower imaging resolution. Thus, to decrease the influence of the finite extent of reconstructed sub-FOV on imaging resolution, in actual experiments, the selection of the reconstructed area faces a fundamental tradeoff between the loss of the high-frequency diffraction and the practicability of the implementation of the reconstruction algorithm. It should be also noted that when pixel SR algorithm is used to achieve an expected sub-pixel resolution, the reconstructed area should be larger than theoretical one calculated by Eq. \ref{ROIL} to guarantee that such a resolution is theoretically achievable.

\subsubsection{Comprehensive influence of multiple factors on imaging resolution}

Based on the above-mentioned analysis, the comprehensive absorption and phase transfer functions of all above-mentioned factors can be denoted as $ATF\left( {\mathbf{u}} \right) = ATFP \cdot TCTF \cdot SCTF \cdot PSTF \cdot RRTF$ and $PTF\left( {\mathbf{u}} \right) = PTFP \cdot TCTF \cdot SCTF \cdot PSTF \cdot RRTF$. Although the frequency response of each transfer function may slightly overshoot for the frequencies exceeds each first cut-off frequency, their contribution to imaging resolution can be neglected because the final imaging resolution is codetermined by multiple parameters, and the overall response value for these high frequencies in $ATF\left( {\mathbf{u}} \right)$ and $PTF\left( {\mathbf{u}} \right)$ after multiplication of each transfer functions will be quite small. Therefore, the final imaging resolution limit is determined by the minimum of the first cut-off frequencies of these sub-transfer functions. For a given LFOCDHM system where each system parameters are determined, we can calculate the resolution limit governed by each transfer function, Eqs. (\ref{ResolutionT}, \ref{ResolutionS}, \ref{ResolutionP}, \ref{ResolutionROI}), and then compare them with ideal coherent diffraction limit $\lambda /2$ to choose the maximal one as the ultimate theoretical imaging resolution. Note that the pixel SR methods are not considered in above analysis. When the SR methods are considered, the theoretical limit resolution will be determined by the maximal value among Eqs. (\ref{ResolutionT}, \ref{ResolutionS}, \ref{ResolutionROI}), and the effective pixel size $\Delta p/w$, $\lambda /2$. In this work, we only consider the cases when no pixel SR methods are employed. The results can be easily extended to the cases when pixel SR methods are involved.

For example, considering the situation in the experiments, the sample-to-senor distance is usually $450 \mu m$, and the source-to-sample distance is about $10 cm$. In addition, the illumination source has central wavelength $600 nm$ with the spectral width $10 nm$ and ${100^2}\pi \mu {m^2}$ luminous area, and the sensor has the pixel size of $1.67 \mu m$ and imaging area of $6466 \times 4615\mu {m^2}$. According to Eqs. (\ref{ResolutionT}, \ref{ResolutionS}, \ref{ResolutionP},\ref{ResolutionROI}), we can find that when no pixel SR methods are employed, the final resolution will be limited by the pixel size. The reconstructed results will be constrained principally by the spectral width $\Delta \lambda$ when the pixel SR methods are adopted. Thus, in a conventional experimental system, the pixel size is the key limiting factor for the high-resolution object reconstruction, but the developed pixel SR methods can effectively solve this spatial resolution reduction problem. In addition, the spectral width of the source is usually another main limiting factor for the resolution improvement, which is difficult to be solved or alleviated only with the numerical methods.

\subsubsection{Optimization of the imaging resolution for a LFOCDHM system}
Our theoretical models can also be utilized to optimize the optical design to improve the imaging resolution when designing a LFOCDHM system. It is recommended that the following procedure should be adopted.

\noindent \textbf{\emph{During the system construction stage:}}

\noindent 1. Choose the light source with the best possible temporal and spatial coherence;

\noindent 2. For low temporal coherent source such as LED, a narrow band-pass filter can be used to increase the temporal coherence of the source;

\noindent 3. For low spatial coherent source with a large light-emitting area, a small pin-hole can be inserted in front of the source to increase the spatial coherence of the source;

\noindent 4. Use an imaging sensor with the smallest possible pixel size to reduce aliasing.

\noindent \textbf{\emph{During the data acquisition stage:}}

\noindent 1. Minimize the sample-to-sensor distance $z_2$ to reduce the influence of temporal coherence of the source;

\noindent 2. Maximum the ratio between source-to-sample distance $z_1$ and sample-to-sensor distance $z_2$ to reduce the influence of spatial coherence of the source;

\noindent 3. Minimize the sample-to-sensor distance $z_2$ to reduce the influence of the finite extent of reconstructed sub-FOV;

\noindent 4. For imaging phase object, use the multi-height phase retrieval algorithm with large sample-to-sensor distances $z_2$ to guarantee reliably phase recovery, especially for low-frequency components.

\noindent It should be emphasized that $z_1$ can only affect the spatial coherence, while $z_2$ can affect the selection of the size of the reconstructed region, the temporal and spatial coherence.

\noindent \textbf{\emph{During the data processing stage:}}

\noindent 1. Choose the largest possible reconstructed sub-FOV to reduce the influence of the finite extent of reconstructed sub-FOV.

\noindent 2. Choose the reconstructed sub-FOV to make the targeted object in the center.

\section{Experiments}
\subsection{Experimental setup}
\begin{figure}[!b]
\centering
\includegraphics[width=\linewidth]{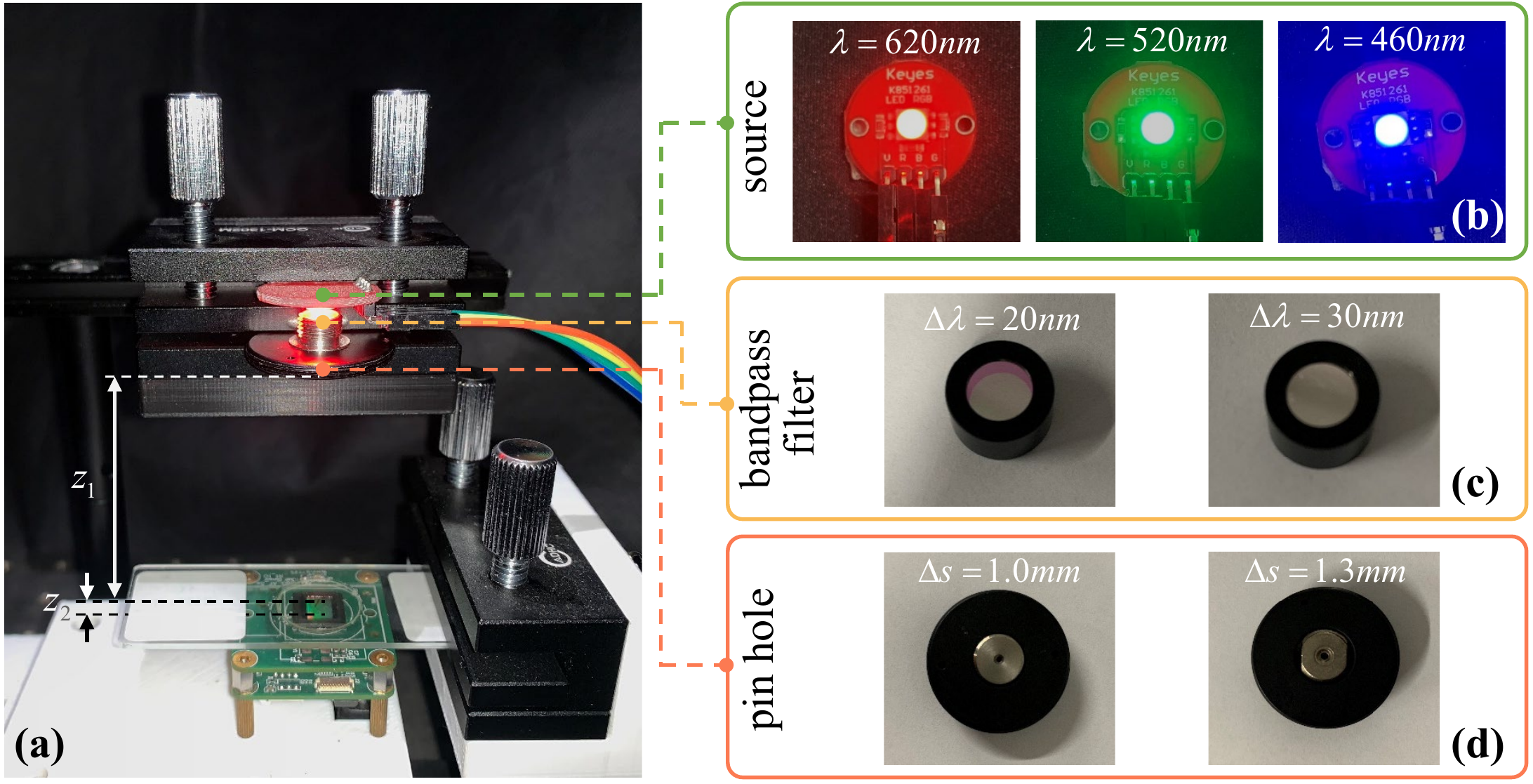}
\caption{(a) The photography of the LFOCDHM system. (b) Three central wavelengths of a light-emitting diode (LED). (c) Two narrow bandpass filters with spectral width $\Delta \lambda  = 20nm$ and $\Delta \lambda  = 30nm$. (d) Two pin-holes with aperture size $\Delta s = 1.0 mm$ and $\Delta s  = 1.3 mm$.}
\label{fig:fig10}
\end{figure}
Figure. \ref{fig:fig10}(a) shows the fundamental experimental system structure. A broadband source (K851261, Keyes, China) providing the different central wavelengths [Fig. \ref{fig:fig10}(b)], illuminates a sample that is mounted on a slide holder, and a CMOS image sensor chip (DMM 27UJ003-ML, the imaging source, Germany) is placed below the sample. To quantify the effect of the above-mentioned factors on the reconstruction results, we will respectively change the temporal [Fig. \ref{fig:fig10}(c)], spatial [Fig. \ref{fig:fig10}(d)] coherence of the light source, the pixel size of the imaging sensor, and the reconstructed region.

\subsection{Influence of temporal coherence on imaging resolution}

\begin{figure}[!b]
\centering
\includegraphics[width=0.8\linewidth]{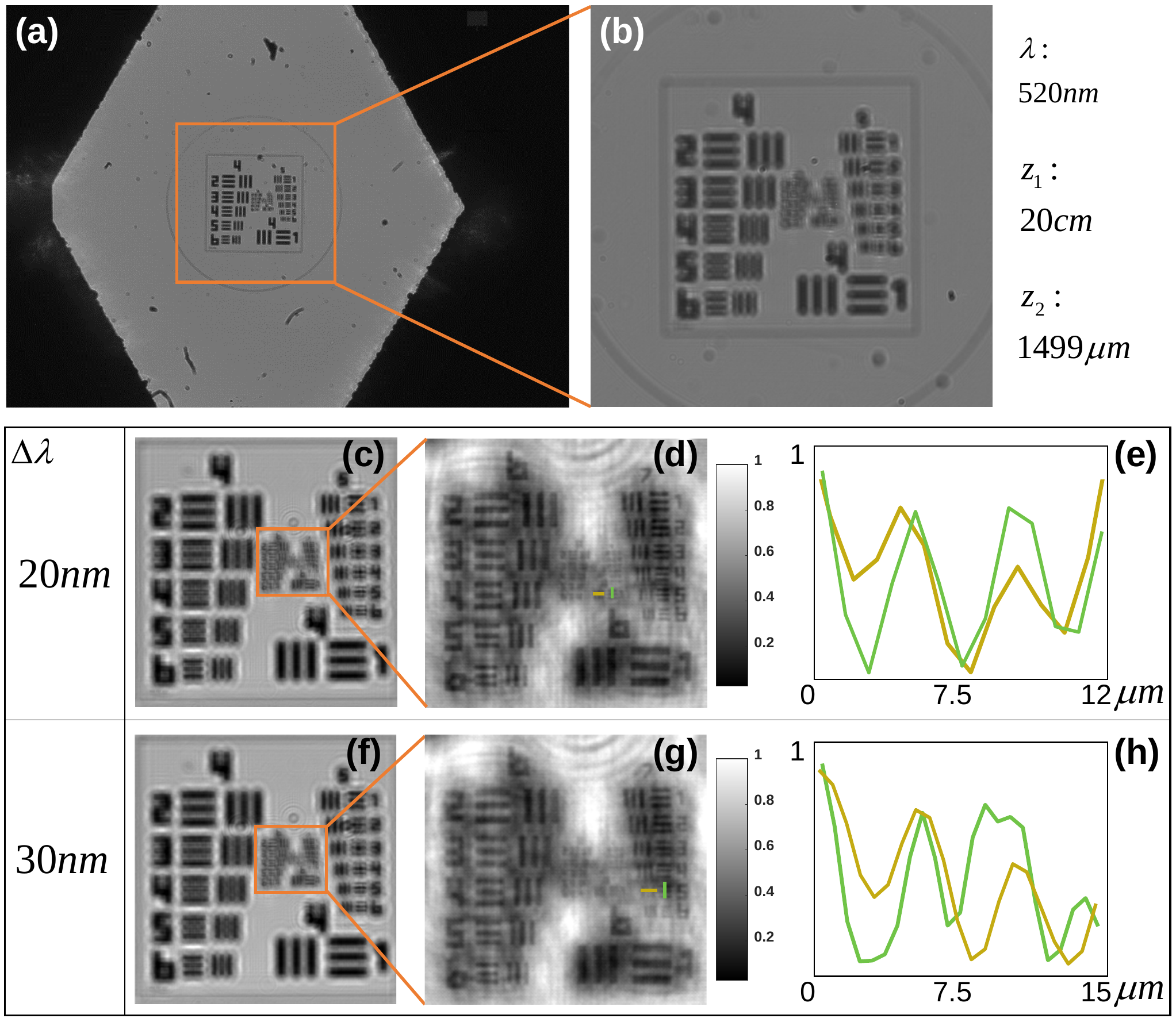}
\caption{The effect of temporal coherence on the spatial resolution. (a) the directly captured image, (b) the region to be reconstructed, the directly reconstructed results with the spectral width $\Delta \lambda  = 20nm$ (c-e) and $\Delta \lambda  = 30nm$ (f-h).}
\label{fig:fig11}
\end{figure}
To quantify the spatial resolution alternation due to the above-mentioned factors respectively, we firstly change the temporal coherence of the light source by introducing different optical band-pass filters (spectral bandwidths $\Delta \lambda  = 20,30nm$) into the experimental system. The partially coherent illumination is provided through a light-emitting diode (LED) which is placed far away (${z_1} \gg 20cm$) from the sample plane to eliminate the effect of the spatial coherence. Figure. \ref{fig:fig11}(a) shows that the raw image directly captured by the camera, and Fig. \ref{fig:fig11}(b) is the reconstructed region which is large enough to avoid its effect on the spatial resolution. The central wavelength of the illumination source is $\sim 520nm$, and the resolution target is $\sim 1499 \mu m$($z_2$) away from the sensor. When the spectral width is $20nm$, the theoretical half-pitch resolution calculated according to Eq. \ref{ResolutionT} is  $1.936 \mu m$, and the actual reconstruction resolution is  $\sim 1.953 \mu m$, as shown in Figs. \ref{fig:fig11}(c,d,e) which corresponds to the 1st element in group $8$ of the resolution target. Similarly, Figs. \ref{fig:fig11}(f,g,h) show that the reconstruction resolution is about $2.461 \mu m$ (5th element in group 7) with the spectral width $\Delta \lambda  = 30nm$, while the theoretical resolution is around $2.371 \mu m$ which lies between the 5th element and 6th element in group 7. Thus, the reconstructed results match well with the theoretical value calculated by Eq. \ref{ResolutionT}. Note that in our experiment, we directly back propagate the image from the sensor plane to the object plane with the angular spectrum method, and no phase retrieval procedure is used to eliminate the twin-image artifacts in the background of the reconstructed images.

\subsection{Influence of spatial coherence on imaging resolution}

\begin{figure}[!htb]
\centering
\includegraphics[width=\linewidth]{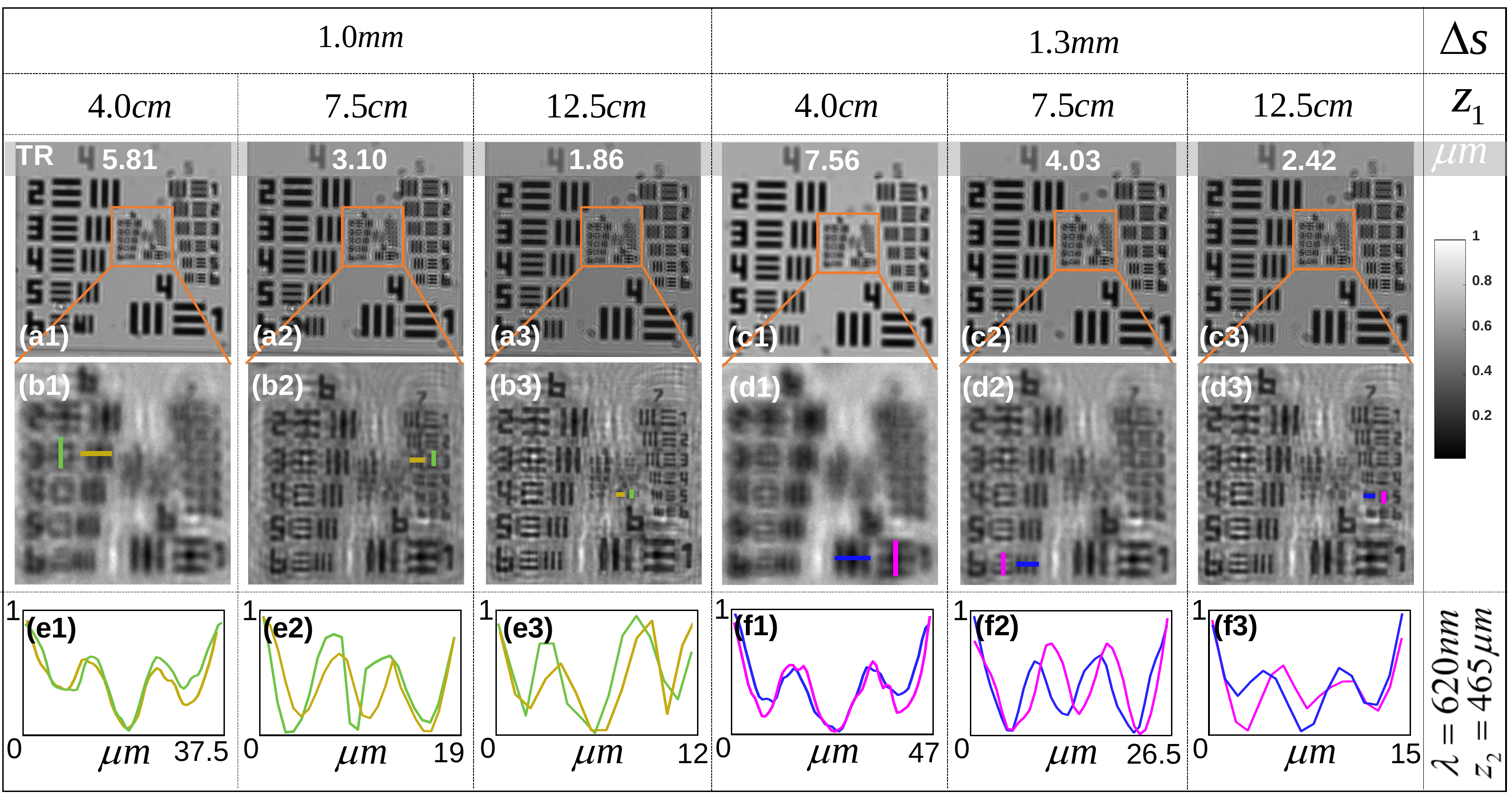}
\caption{The effect of spatial coherence on the spatial resolution. `TR' is the abbreviation of theoretical resolution. (a1-b3) The reconstructed results with $\Delta s = 1mm$. (c1-d3) The reconstructed results with $\Delta s = 1.3mm$. (e1-f3) The line profiles of the corresponding marks in (b1-b3,d1-d3).}
\label{fig:fig12}
\end{figure}

Next, we change the spatial coherence of the source by inserting the different pin-holes (the diameter of the pin-holes $\Delta s = 1.0, 1.3mm$) to verify the correctness of Eq. \ref{ResolutionS}. The luminous area of a LED is usually in the several hundreds of microns order of magnitude, thus  in order to show the influence of spatial coherence on resolution more intuitively, a diffuser is placed between the source and pin-hole to ensure that the luminous area is the size of the pin hole. The center wavelength $\lambda $ is $\sim 620 nm$ and the sample-to-sensor distance is $z_{2} = 465 \mu m$. Figure \ref{fig:fig12} shows the reconstruction results which are recovered by back-propagating the captured image to the object plane with angular spectrum method. When $\Delta s = 1 mm$, the reconstructed results with different the source-to-sample distances $z_{1}$  are shown in Figs. \ref{fig:fig12}(b1-b3). When $z_1$ is $4 cm$, the theoretical half-pitch resolution is $5.81 \mu m$, and the actual reconstructed result is $\sim 6.20 \mu m$, corresponding to the 3rd element of group 6. Since the 4th element in group 6 corresponds to the half-pitch resolution of $5.52 \mu m$, it can hardly be distinguished, as shown in Fig. \ref{fig:fig12}(b1). In addition, when $\Delta s = 1.3 mm$, the experimental results are also agreed well with the theoretical values, as shown in Figs. \ref{fig:fig12}(d1-d3). The line profiles along different resolution elements are respectively illustrated in Figs. \ref{fig:fig12}(f1-f3). On the other hand, when $z_{1}$ is fixed, a smaller $\Delta s$ provides higher resolution. Thus, in the actual experiments, we can simply increase the source-to-sample distance $z_1$ to reduce the influence of spatial coherence, which is equivalent to reducing $\Delta s$.
\subsection{Influence of pixel size on imaging resolution}

\begin{figure}[!htb]
\centering
\includegraphics[width= 0.9\linewidth]{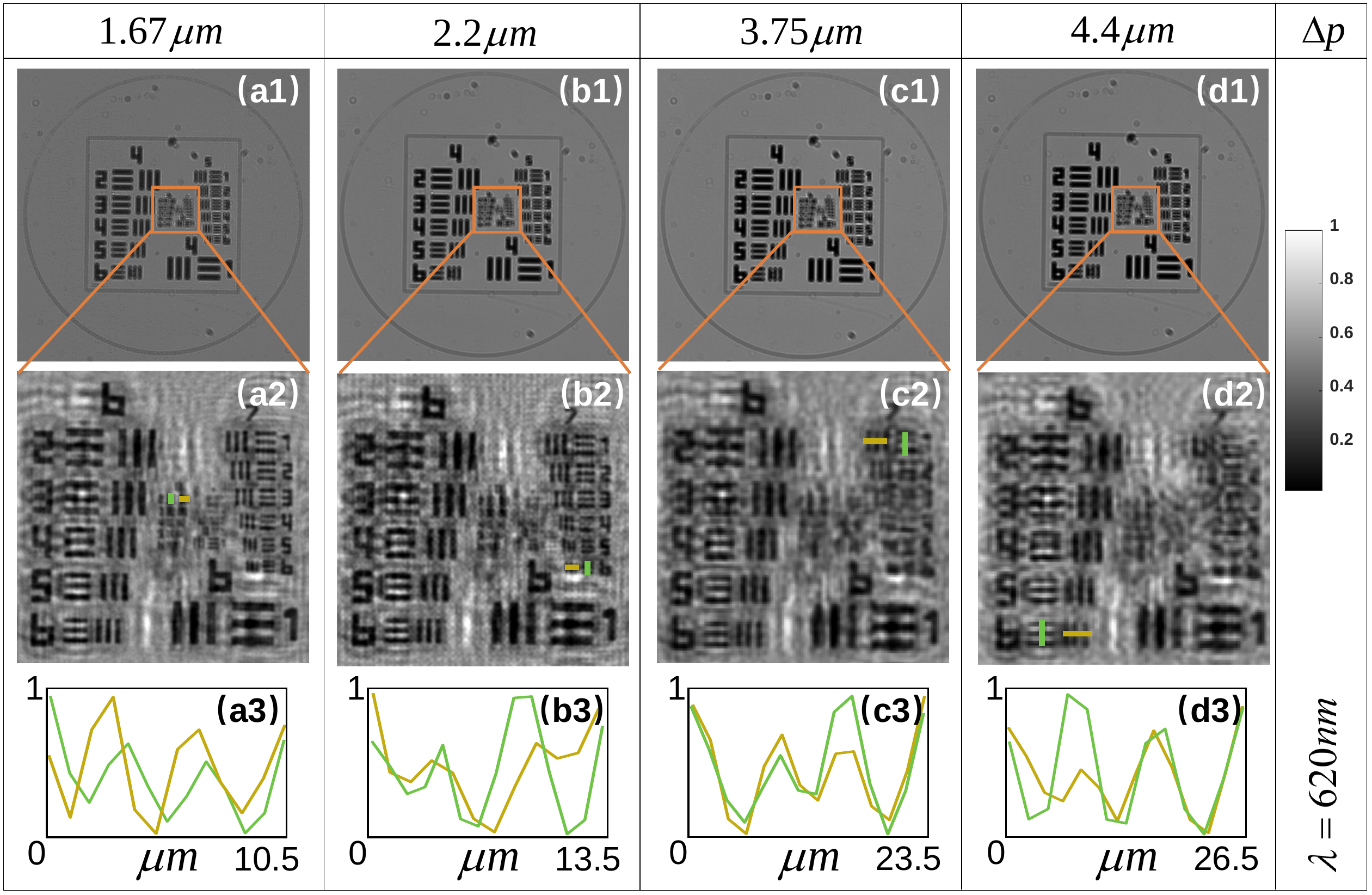}
\caption{The effect of pixel size on the spatial resolution. The directly reconstructed results with different pixel sizes $1.67 \mu m$ (a1-a3), $2.2 \mu m$ (b1-b3), $3.75 \mu m$ (c1-c3), $4.4 \mu m$ (d1-d3).}
\label{fig:fig13}
\end{figure}

In actual experiments, the pixel size of the image sensor is a key factor directly limiting the achievable spatial resolution. Although increasing the pixel resolution and reducing the pixel size has already become the major trend in consumer electronics, the minimum pixel size of the commercially available imaging sensor is around $0.8 \mu m$, which is much larger than the coherent diffraction resolution limit. In order to give an intuitive comparison of the influence of pixel size on imaging resolution, we use the cameras with the different pixel sizes ($1.67 \mu m$, $2.2 \mu m$, $3.75 \mu m$, $4.4 \mu m$) to record the diffraction patterns. Figure \ref{fig:fig13}(a1-d1) show the reconstructed area, and the reconstructed results are illustrated in Figs. \ref{fig:fig13}(a2-d2). The wavelength of source used in the system is $620 nm$ while the source-to-sample distance $z_1$ is large enough  (usually ${z_1} \gg 20cm$) to exclude the influence of spatial coherence, and the sample-to-sensor distance $z_2$ is $465 \mu m$. The line profiles corresponding to the smallest resolvable elements are shown in Figs. \ref{fig:fig13}(a3-d3), suggesting that the experimental results are in agreement with the theoretical values limited by pixel sizes.

\subsection{Influence of the reconstructed region on imaging resolution}

\begin{figure}[!htb]
\centering
\includegraphics[width=\linewidth]{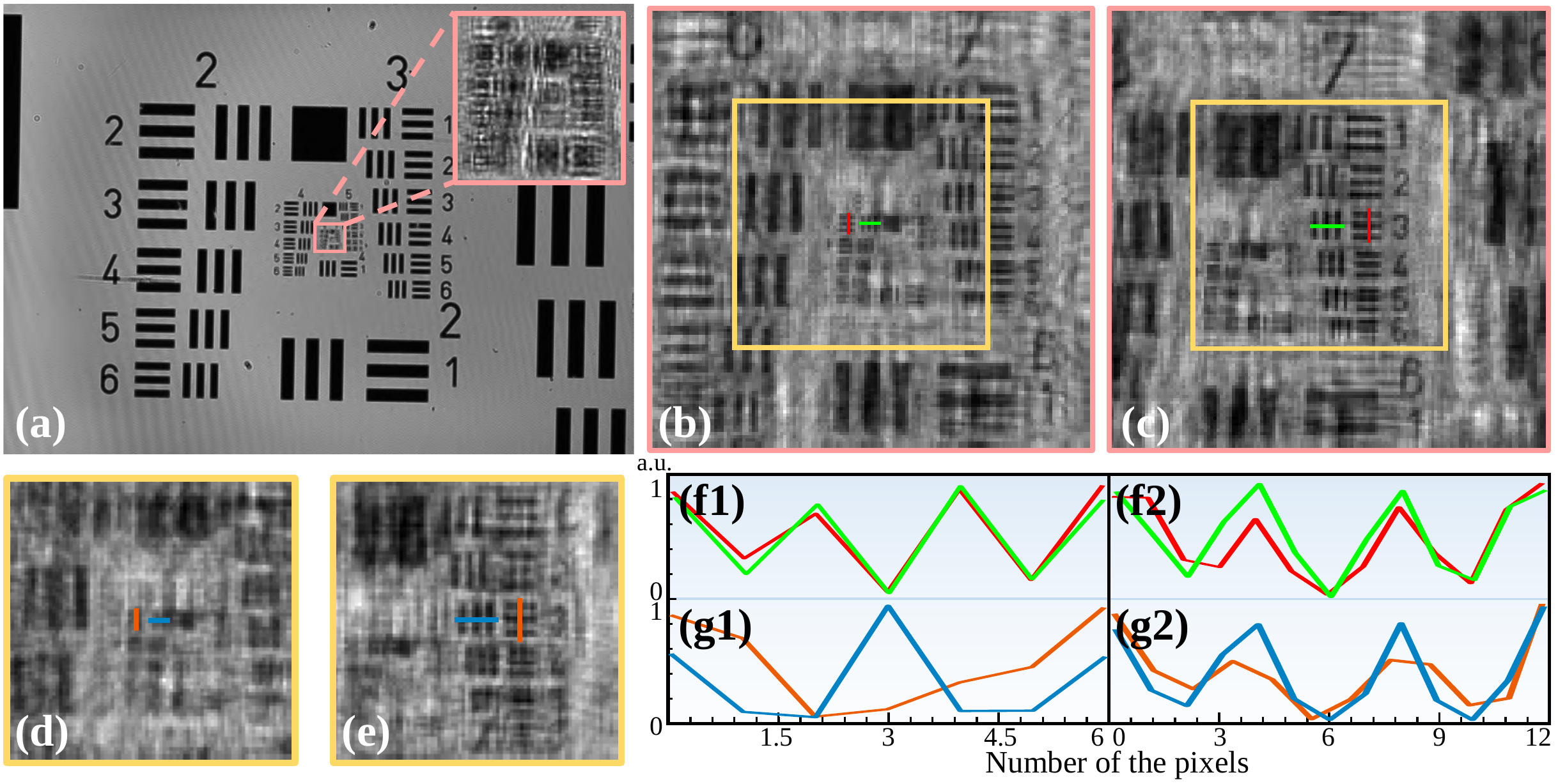}
\caption{The effect of reconstructed region on the spatial resolution. (a) is the raw image and the reconstructed region size of (b-c) corresponds to the pink rectangular area in (a). (d-e) is the directly reconstructed results with the different recovered areas separately corresponding to the yellow rectangular area in (b-c). (f1-f2), (g1-g2) are the line profiles separately corresponding to (b-c), (d-e).}
\label{fig:fig14}
\end{figure}

In this experiment, the center wavelength of the light source $\lambda$ is $620 nm$, and the sample-to-sensor distance $z_2$ is $547 \mu m$. According to Eq. \ref{ResolutionROI}, the size of the selected area for the reconstruction will affect the final imaging resolution. Figure \ref{fig:fig14}(a) gives the whole captured image, and the pink rectangular area (length of side $198 \mu m$) was extracted for the holographic reconstruction. The result is shown in Fig. \ref{fig:fig14}(b), and corresponding line profiles are shown in Fig. \ref{fig:fig14}(f1), suggesting that the resolution is at least $1.74 \mu m$. When we select another region nearby with the same size, we can obtain the reconstruction result shown Fig. \ref{fig:fig14}(c). If we reduce the size of the reconstructed region to the the yellow boxed area (length of side $110 \mu m$) in Figs. \ref{fig:fig14}(b-c), the results shown in Figs. \ref{fig:fig14}(d-e) indicate that the reconstructed resolution will decrease significantly. The line profiles in Figs. \ref{fig:fig14}(g1-g2) manifest that the resolution is reduced to only $3.10 \mu m$ (3rd element in group 7), which is again in accordance with the theoretical prediction.

In addition to the size of reconstructed sub-FOV, the location of the object to be measured in the selected reconstructed sub-FOV will also affect the reconstructed resolution. As shown in Figs. \ref{fig:fig14}(b-c), we can find that the 2rd element in group 8 can be distinguishable in Fig. \ref{fig:fig14}(b) but not in Fig. \ref{fig:fig14}(c). Thus, in order to ensure the expected high reconstruction resolution, the reconstructed sub-FOV should not be too small and the objects to be reconstructed are supposed to be close to the limited central region for each reconstructed sub-FOV. Meanwhile, the object-to-sensor distance $z_2$ should not be too large according to Eq. \ref{ResolutionROI}. Otherwise, the reconstructed region needs to be expanded accordingly to ensure the reconstruction resolution, which may significantly prolong the processing time and create difficulties in practical implementation of the reconstruction algorithm.

\section{Conclusions and Discussions}

In this work, we have conducted a systematical research on the effect of five major factors on imaging resolution of a LFOCDHM system, i.e., the sample-to-sensor distance, spatial and temporal coherence of the illumination, finite size of the equally spaced sensor pixels, and finite extent of the image sub-FOV used for the reconstruction. From the above analysis and experiments, it can be deduced that the most limiting factor restricting the imaging resolution of LFOCDHM is the sensor pixel size because the side-effect arising from other experimental factors is relatively easy to handle. For example, using a laser as an ideal temporally coherent light source, increasing source-to-sample distance to obtain a close to the ideal spatially coherent source. To reduce the effective size of the imaging sensor, pixel SR algorithms should be used. But even so, using an imaging sensor with smaller pixel size can still improve the quality of the SR reconstructions. Specifically, assuming that the expected resolution to be reconstructed is around $1 \mu m$, and the up-sampling factor $w$ will be different for various pixel sizes. When the pixel size is much closer to the desired resolution, the $w$ will be smaller, so less information for the reconstruction is required. When a higher up-sampling factor $w$ is required (for large pixel size), more criss-crossed frequency gaps will appear, which can never be recovered even pixel SR reconstruction algorithms are used. Thus, for LFOCDHM techniques, a smaller pixel size is very helpfully to achieve higher resolution and need less information to reach the expected super-resolved resolution. On the other hand, using LED as the light source can make the system more compact, portable, low-cost. But the coherence length of the LED will also affect the reconstructed resolution. According to Eqs. (\ref{ResolutionT},\ref{ResolutionS}), increasing ${z_1}$ and decreasing ${z_2}$ can effectively improve the coherence of light sources and improve the imaging resolution. Furthermore, decreasing ${z_2}$ can reduce the reconstructed area according to Eq. \ref{ROIL} when the desired resolution is determined.

The analysis of these parameters based on transfer functions has given the quantitative resolution limit determined by the minimum first cut-off frequency of these transfer functions. According to the quantitative relationship, the preliminary estimates of the ultimate resolution are available after employing the SR methods. Thus, the derived theoretical models can provide useful guidance to choosing the appropriate system parameters to obtain higher imaging resolution. To verify the validity of each theoretical model, we have used the variable-controlling method and only changed only one or two parameters during each experiment. The resolution target has been used to quantify the imaging resolution. The experimental results have confirmed the validity of our theoretical models.

Finally, it should be mentioned that, although in this work we have demonstrated how our theoretical models can be utilized to improve the imaging resolution by optimizing the optical design of a LFOCDHM system, it should also be possible to counteract the effects of these imperfect system parameters through certain computational approaches. Based on the transfer functions we have derived, we can easily establish the forward image formation model (from object to image) for a given LFOCDHM system. Then certain mathematical algorithm should be adopted to recover the ideal object information from the actual measurement, i.e., to solve the corresponding inverse problem. In future work, we will make effort to address the resolution reduction associated with these factors and compensate for their adverse impact through post-processing algorithms.

\section*{Funding}
This work was supported by the National Natural Science Foundation of China (61722506, 11574152), Final Assembly ``13th Five-Year Plan" Advanced Research Project of China (30102070102), Equipment Advanced Research Fund of China (61404150202), National Defense Science and Technology Foundation of China (0106173), Outstanding Youth Foundation of Jiangsu Province (BK20170034), The Key Research and Development Program of Jiangsu Province (BE2017162), ``333 Engineering" Research Project of Jiangsu Province (BRA2016407), Fundamental Research Funds for the Central Universities (30917011204).
\bibliography{OSA-template}

\begin{thebibliography}{10}
\newcommand{\enquote}[1]{``#1''}

\bibitem{maricq1973patterns}
H.~R. Maricq and E.~C. LeRoy, \enquote{Patterns of finger capillary
  abnormalities in connective tissue disease by ``wide-field'' microscopy,}
  {\protect\JournalTitle{Arthritis \& Rheumatism}} \textbf{16}, 619--628
  (1973).

\bibitem{Huisman2010Creation}
A.~Huisman, A.~Looijen, S.~M. V.~D. Brink, and P.~J.~V. Diest,
  \enquote{Creation of a fully digital pathology slide archive by high-volume
  tissue slide scanning,} {\protect\JournalTitle{Human Pathology}} \textbf{41},
  751--757 (2010).

\bibitem{ma2007use}
B.~Ma, T.~Zimmermann, M.~Rohde, S.~Winkelbach, F.~He, W.~Lindenmaier, and K.~E.
  Dittmar, \enquote{Use of autostitch for automatic stitching of microscope
  images,} {\protect\JournalTitle{Micron}} \textbf{38}, 492--499 (2007).

\bibitem{mico2006synthetic}
V.~Mico, Z.~Zalevsky, P.~Garc{\'\i}a-Mart{\'\i}nez, and J.~Garc{\'\i}a,
  \enquote{Synthetic aperture superresolution with multiple off-axis
  holograms,} {\protect\JournalTitle{Journal of the Optical Society of America
  A}} \textbf{23}, 3162--3170 (2006).

\bibitem{yuan2008angular}
C.~Yuan, H.~Zhai, and H.~Liu, \enquote{Angular multiplexing in pulsed digital
  holography for aperture synthesis,} {\protect\JournalTitle{Optics Letters}}
  \textbf{33}, 2356--2358 (2008).

\bibitem{Hillman2009High}
T.~R. Hillman, G.~Thomas, S.~A. Alexandrov, and D.~D. Sampson,
  \enquote{High-resolution, wide-field object reconstruction with synthetic
  aperture fourier holographic optical microscopy,}
  {\protect\JournalTitle{Optics Express}} \textbf{17}, 7873--7892 (2009).

\bibitem{kim2014common}
Y.~Kim, H.~Shim, K.~Kim, H.~Park, J.~H. Heo, J.~Yoon, C.~Choi, S.~Jang, and
  Y.~Park, \enquote{Common-path diffraction optical tomography for
  investigation of three-dimensional structures and dynamics of biological
  cells,} {\protect\JournalTitle{Optics Express}} \textbf{22}, 10398--10407
  (2014).

\bibitem{kim2014profiling}
Y.~Kim, H.~Shim, K.~Kim, H.~Park, S.~Jang, and Y.~Park, \enquote{Profiling
  individual human red blood cells using common-path diffraction optical
  tomography,} {\protect\JournalTitle{Scientific Reports}} \textbf{4}, 6659
  (2014).

\bibitem{lim2015comparative}
J.~Lim, K.~Lee, K.~H. Jin, S.~Shin, S.~Lee, Y.~Park, and J.~C. Ye,
  \enquote{Comparative study of iterative reconstruction algorithms for missing
  cone problems in optical diffraction tomography,}
  {\protect\JournalTitle{Optics Express}} \textbf{23}, 16933--16948 (2015).

\bibitem{Zheng2013Wide}
G.~Zheng, R.~Horstmeyer, and C.~Yang, \enquote{Wide-field, high-resolution
  fourier ptychographic microscopy,} {\protect\JournalTitle{Nature Photonics}}
  \textbf{7}, 739--745 (2013).

\bibitem{ou2013quantitative}
X.~Ou, R.~Horstmeyer, C.~Yang, and G.~Zheng, \enquote{Quantitative phase
  imaging via fourier ptychographic microscopy,} {\protect\JournalTitle{Optics
  Letters}} \textbf{38}, 4845--4848 (2013).

\bibitem{tian2014multiplexed}
L.~Tian, X.~Li, K.~Ramchandran, and L.~Waller, \enquote{Multiplexed coded
  illumination for fourier ptychography with an led array microscope,}
  {\protect\JournalTitle{Biomedical Optics Express}} \textbf{5}, 2376--2389
  (2014).

\bibitem{ou2015high}
X.~Ou, R.~Horstmeyer, G.~Zheng, and C.~Yang, \enquote{High numerical aperture
  fourier ptychography: principle, implementation and characterization,}
  {\protect\JournalTitle{Optics Express}} \textbf{23}, 3472--3491 (2015).

\bibitem{zuo2016adaptive}
C.~Zuo, J.~Sun, and Q.~Chen, \enquote{Adaptive step-size strategy for
  noise-robust fourier ptychographic microscopy,} {\protect\JournalTitle{Optics
  Express}} \textbf{24}, 20724--20744 (2016).

\bibitem{sun2016efficient}
J.~Sun, Q.~Chen, Y.~Zhang, and C.~Zuo, \enquote{Efficient positional
  misalignment correction method for fourier ptychographic microscopy,}
  {\protect\JournalTitle{Biomedical Optics Express}} \textbf{7}, 1336--1350
  (2016).

\bibitem{sun2016sampling}
J.~Sun, Q.~Chen, Y.~Zhang, and C.~Zuo, \enquote{Sampling criteria for fourier
  ptychographic microscopy in object space and frequency space,}
  {\protect\JournalTitle{Optics Express}} \textbf{24}, 15765--15781 (2016).

\bibitem{zheng2011epetri}
G.~Zheng, S.~A. Lee, Y.~Antebi, M.~B. Elowitz, and C.~Yang, \enquote{The epetri
  dish, an on-chip cell imaging platform based on subpixel perspective sweeping
  microscopy (spsm),} {\protect\JournalTitle{Proceedings of the National
  Academy of Sciences}} \textbf{108}, 16889--16894 (2011).

\bibitem{luo2016propagation}
W.~Luo, Y.~Zhang, Z.~G{\"o}r{\"o}cs, A.~Feizi, and A.~Ozcan,
  \enquote{Propagation phasor approach for holographic image reconstruction,}
  {\protect\JournalTitle{Scientific Reports}} \textbf{6}, 22738 (2016).

\bibitem{rivenson2018deep}
Y.~Rivenson, H.~Ceylan~Koydemir, H.~Wang, Z.~Wei, Z.~Ren, H.~G{\"u}nayd{\i}n,
  Y.~Zhang, Z.~G{\"o}r{\"o}cs, K.~Liang, D.~Tseng \emph{et~al.}, \enquote{Deep
  learning enhanced mobile-phone microscopy,} {\protect\JournalTitle{ACS
  Photonics}} \textbf{5}, 2354--2364 (2018).

\bibitem{zhang2018lensfree}
J.~Zhang, Q.~Chen, J.~Li, J.~Sun, and C.~Zuo, \enquote{Lensfree dynamic
  super-resolved phase imaging based on active micro-scanning,}
  {\protect\JournalTitle{Optics Letters}} \textbf{43}, 3714--3717 (2018).

\bibitem{garcia2006immersion}
J.~Garcia-Sucerquia, W.~Xu, M.~Jericho, and H.~J. Kreuzer, \enquote{Immersion
  digital in-line holographic microscopy,} {\protect\JournalTitle{Optics
  Letters}} \textbf{31}, 1211--1213 (2006).

\bibitem{ozcan2016lensless}
A.~Ozcan and E.~McLeod, \enquote{Lensless imaging and sensing,}
  {\protect\JournalTitle{Annual Review of Biomedical Engineering}} \textbf{18},
  77--102 (2016).

\bibitem{mudanyali2010compact}
O.~Mudanyali, D.~Tseng, C.~Oh, S.~O. Isikman, I.~Sencan, W.~Bishara,
  C.~Oztoprak, S.~Seo, B.~Khademhosseini, and A.~Ozcan, \enquote{Compact,
  light-weight and cost-effective microscope based on lensless incoherent
  holography for telemedicine applications,} {\protect\JournalTitle{Lab on a
  Chip}} \textbf{10}, 1417--1428 (2010).

\bibitem{su2010compact}
T.-W. Su, A.~Erlinger, D.~Tseng, and A.~Ozcan, \enquote{Compact and
  light-weight automated semen analysis platform using lensfree on-chip
  microscopy,} {\protect\JournalTitle{Analytical Chemistry}} \textbf{82},
  8307--8312 (2010).

\bibitem{cui2008lensless}
X.~Cui, L.~M. Lee, X.~Heng, W.~Zhong, P.~W. Sternberg, D.~Psaltis, and C.~Yang,
  \enquote{Lensless high-resolution on-chip optofluidic microscopes for
  caenorhabditis elegans and cell imaging,} {\protect\JournalTitle{Proceedings
  of the National Academy of Sciences}} \textbf{105}, 10670--10675 (2008).

\bibitem{bishara2011holographic}
W.~Bishara, U.~Sikora, O.~Mudanyali, T.-W. Su, O.~Yaglidere, S.~Luckhart, and
  A.~Ozcan, \enquote{Holographic pixel super-resolution in portable lensless
  on-chip microscopy using a fiber-optic array,} {\protect\JournalTitle{Lab on
  a Chip}} \textbf{11}, 1276--1279 (2011).

\bibitem{greenbaum2012imaging}
A.~Greenbaum, W.~Luo, T.-W. Su, Z.~G{\"o}r{\"o}cs, L.~Xue, S.~O. Isikman, A.~F.
  Coskun, O.~Mudanyali, and A.~Ozcan, \enquote{Imaging without lenses:
  achievements and remaining challenges of wide-field on-chip microscopy,}
  {\protect\JournalTitle{Nature Methods}} \textbf{9}, 889 (2012).

\bibitem{barton1991removing}
J.~Barton, \enquote{Removing multiple scattering and twin images from
  holographic images,} {\protect\JournalTitle{Physical Review Letters}}
  \textbf{67}, 3106 (1991).

\bibitem{latychevskaia2007solution}
T.~Latychevskaia and H.-W. Fink, \enquote{Solution to the twin image problem in
  holography,} {\protect\JournalTitle{Physical Review Letters}} \textbf{98},
  233901 (2007).

\bibitem{park2003super}
S.~C. Park, M.~K. Park, and M.~G. Kang, \enquote{Super-resolution image
  reconstruction: a technical overview,} {\protect\JournalTitle{IEEE Signal
  Processing Magazine}} \textbf{20}, 21--36 (2003).

\bibitem{zhang2017adaptive}
J.~Zhang, J.~Sun, Q.~Chen, J.~Li, and C.~Zuo, \enquote{Adaptive
  pixel-super-resolved lensfree in-line digital holography for wide-field
  on-chip microscopy,} {\protect\JournalTitle{Scientific Reports}} \textbf{7},
  11777 (2017).

\bibitem{parrent1965resolution}
G.~B. Parrent and G.~O. Reynolds, \enquote{Resolution limitations of lensless
  photography,} {\protect\JournalTitle{Optical Engineering}} \textbf{3}, 306219
  (1965).

\bibitem{agbana2017aliasing}
T.~E. Agbana, H.~Gong, A.~S. Amoah, V.~Bezzubik, M.~Verhaegen, and G.~Vdovin,
  \enquote{Aliasing, coherence, and resolution in a lensless holographic
  microscope,} {\protect\JournalTitle{Optics Letters}} \textbf{42}, 2271--2274
  (2017).

\bibitem{xu2005imaging}
L.~Xu, X.~Peng, Z.~Guo, J.~Miao, and A.~Asundi, \enquote{Imaging analysis of
  digital holography,} {\protect\JournalTitle{Optics Express}} \textbf{13},
  2444--2452 (2005).

\bibitem{kelly2009resolution}
D.~P. Kelly, B.~M. Hennelly, N.~Pandey, T.~J. Naughton, and W.~T. Rhodes,
  \enquote{Resolution limits in practical digital holographic systems,}
  {\protect\JournalTitle{Optical Engineering}} \textbf{48}, 095801 (2009).

\bibitem{hao2011resolution}
Y.~Hao and A.~Asundi, \enquote{Resolution analysis of a digital holography
  system,} {\protect\JournalTitle{Applied Optics}} \textbf{50}, 183--193
  (2011).

\bibitem{doblas2015study}
A.~Doblas, E.~S{\'a}nchez-Ortiga, M.~Mart{\'\i}nez-Corral, and
  J.~Garcia-Sucerquia, \enquote{Study of spatial lateral resolution in off-axis
  digital holographic microscopy,} {\protect\JournalTitle{Optics
  Communications}} \textbf{352}, 63--69 (2015).

\bibitem{kelly2013filtering}
D.~P. Kelly and D.~Claus, \enquote{Filtering role of the sensor pixel in
  fourier and fresnel digital holography,} {\protect\JournalTitle{Applied
  Optics}} \textbf{52}, A336--A345 (2013).

\bibitem{ozcan2008ultra}
A.~Ozcan and U.~Demirci, \enquote{Ultra wide-field lens-free monitoring of
  cells on-chip,} {\protect\JournalTitle{Lab on a Chip}} \textbf{8}, 98--106
  (2008).

\bibitem{luo2016pixel}
W.~Luo, Y.~Zhang, A.~Feizi, Z.~G{\"o}r{\"o}cs, and A.~Ozcan, \enquote{Pixel
  super-resolution using wavelength scanning,} {\protect\JournalTitle{Light:
  Science \& Applications}} \textbf{5}, e16060 (2016).

\bibitem{tseng2010lensfree}
D.~Tseng, O.~Mudanyali, C.~Oztoprak, S.~O. Isikman, I.~Sencan, O.~Yaglidere,
  and A.~Ozcan, \enquote{Lensfree microscopy on a cellphone,}
  {\protect\JournalTitle{Lab on a Chip}} \textbf{10}, 1787--1792 (2010).

\bibitem{kesavan2014high}
S.~V. Kesavan, F.~Momey, O.~Cioni, B.~David-Watine, N.~Dubrulle, S.~Shorte,
  E.~Sulpice, D.~Freida, B.~Chalmond, J.~Dinten \emph{et~al.},
  \enquote{High-throughput monitoring of major cell functions by means of
  lensfree video microscopy,} {\protect\JournalTitle{Scientific Reports}}
  \textbf{4}, 5942 (2014).

\bibitem{ludwig2015calling}
S.~K. Ludwig, C.~Tokarski, S.~N. Lang, L.~A. van Ginkel, H.~Zhu, A.~Ozcan, and
  M.~W. Nielen, \enquote{Calling biomarkers in milk using a protein microarray
  on your smartphone,} {\protect\JournalTitle{PLoS One}} \textbf{10}, e0134360
  (2015).

\bibitem{xiong2018optimized}
Z.~Xiong, J.~E. Melzer, J.~Garan, and E.~McLeod, \enquote{Optimized sensing of
  sparse and small targets using lens-free holographic microscopy,}
  {\protect\JournalTitle{Optics Express}} \textbf{26}, 25676--25692 (2018).

\bibitem{bishara2010lensfree}
W.~Bishara, T.-W. Su, A.~F. Coskun, and A.~Ozcan, \enquote{Lensfree on-chip
  microscopy over a wide field-of-view using pixel super-resolution,}
  {\protect\JournalTitle{Optics Express}} \textbf{18}, 11181--11191 (2010).

\bibitem{kirkland2010advanced}
E.~J. Kirkland, \emph{Advanced computing in electron microscopy} (Springer
  Science \& Business Media, 2010).

\bibitem{zuo2017high}
C.~Zuo, J.~Sun, J.~Li, J.~Zhang, A.~Asundi, and Q.~Chen,
  \enquote{High-resolution transport-of-intensity quantitative phase microscopy
  with annular illumination,} {\protect\JournalTitle{Scientific Reports}}
  \textbf{7}, 7654 (2017).

\bibitem{goodman2005introduction}
J.~W. Goodman, \emph{Introduction to Fourier optics} (Roberts and Company
  Publishers, 2005).

\bibitem{hamilton1984improved}
D.~Hamilton, C.~Sheppard, and T.~Wilson, \enquote{Improved imaging of phase
  gradients in scanning optical microscopy,} {\protect\JournalTitle{Journal of
  Microscopy}} \textbf{135}, 275--286 (1984).

\bibitem{feng_resolution_2017}
S.~Feng and J.~Wu, \enquote{Resolution enhancement method for lensless in-line
  holographic microscope with spatially-extended light source,}
  {\protect\JournalTitle{Optics Express}} \textbf{25}, 24735 (2017).

\bibitem{gonzalez1977digital}
R.~C. Gonzalez and P.~Wintz, \emph{Digital image processing}, 13 (1977).

\bibitem{zheng2010sub}
G.~Zheng, S.~A. Lee, S.~Yang, and C.~Yang, \enquote{Sub-pixel resolving
  optofluidic microscope for on-chip cell imaging,} {\protect\JournalTitle{Lab
  on a Chip}} \textbf{10}, 3125--3129 (2010).

\bibitem{miao1998phase}
J.~Miao, D.~Sayre, and H.~Chapman, \enquote{Phase retrieval from the magnitude
  of the fourier transforms of nonperiodic objects,}
  {\protect\JournalTitle{Journal of the Optical Society of America A}}
  \textbf{15}, 1662--1669 (1998).

\end{thebibliography}

\end{document}